# Social Integration and Housing Behaviours of Immigrants: Evidence from Singapore's Public Housing Market

Yi Fan,[*Ω] Ho Pin Teo,[†] Yong Tu,[‡] Wayne Xinwei Wan[§Ω]

October 2025

**Abstract**

This study investigates the impact of social integration on immigrants' housing behaviours from a temporal perspective, using Singapore's differential public housing policies on immigrants as a quasi-natural experiment. With the support of a local town council, we conducted a survey on social integration among 1,128 immigrant and local households living in public housing estates. In the public open rental housing market—primarily accommodating yet-to-integrate immigrants—we find immigrant renters live up to 3.04% farther from their workplace and pay lower rents up to 0.67% per additional year of residency. Such impacts are more substantial among minority ethnic groups. The results remain robust when using alternative subjective or objective measures of social integration. However, in the public resale housing market—primarily accommodating native and well-integrated naturalised citizens—we find that naturalised citizens face no price premiums relative to native homebuyers, implying no further effect of integration on housing prices after well-integration. This study extends the literature of spatial assimilation focusing on ethnic residential segregations and is generalizable to cities with few ethnic enclaves.

*Department of Real Estate, National University of Singapore. E-mail: yi.fan@nus.edu.sg. Yi Fan acknowledges financial support from NUS Startup Grant No. R-297-000-134-133.

†College of Design and Engineering, National University of Singapore E-mail: hopin.teo@gmail.com.

‡Department of Real Estate, National University of Singapore E-mail: rsttuy@nus.edu.sg Yong Tu acknowledges financial support from MOE Academic Research Fund No. R-297-000-136-115.

§Department of Banking and Finance, Monash University. E-mail: wayne.wan@monash.edu.

Author names are in alphabetical sequence, and Ω indicates the corresponding authors. We thank Yaopei Wang for excellent research support.

# Social Integration and Housing Behaviours of Immigrants:

# Evidence from Singapore's Public Housing Market

**Abstract**

This study investigates the impact of social integration on immigrants' housing behaviours from a temporal perspective, using Singapore's differential public housing policies on immigrants as a quasi-natural experiment. With the support of a local town council, we conducted a survey on social integration among 1,128 immigrant and local households living in public housing estates. In the public open rental housing market—primarily accommodating yet-to-integrate immigrants—we find immigrant renters live up to 3.04% farther from their workplace and pay lower rents up to 0.67% per additional year of residency. Such impacts are more substantial among minority ethnic groups. The results remain robust when using alternative subjective or objective measures of social integration. However, in the public resale housing market—primarily accommodating native and well-integrated naturalised citizens—we find that naturalised citizens face no price premiums relative to native homebuyers, implying no further effect of integration on housing prices after well-integration. This study extends the literature of spatial assimilation focusing on ethnic residential segregations and is generalizable to cities with few ethnic enclaves.

# 1 Introduction

In the past decade, rapid globalization has spurred cross-border migration (Abel & Sander, 2014; Czaika & De Hass, 2014). With increasing migration, many host cities have new socioeconomic problems, such as informal settlements, residential segregation, inferior education, and labour market inequality (van der Laan Bouma-Doff, 2007; Cutler et al., 2008; Boeri et al., 2015; Abramitzky & Boustan, 2017). Since many of these issues have been attributed to immigrants' difficulty in integrating with their migration destinations, social integration of immigrants in host societies has attracted extensive public interests (Bloomberg, 2019; Financial Times, 2019). Some governments have been proactive in helping immigrants integrate into the host society through various means, such as setting up social support associations to mitigate discrimination and providing flexible education pathways for their children (Schmidt Sr., 2007). In contrast, other societies lack sufficient policies to support immigrants' social integration—and, in some extreme cases, treat immigrants in an openly hostile manner (e.g., Nicholls et al., 2020). One possible reason is that the impacts of social integration policies on social sustainability have not been fully realized. Therefore, gaining a better understanding of how social integration influences immigrants' behaviours can help inform policymaking towards social sustainability.

Among the outcomes of social integration, this study focuses on immigrants' housing behaviours, because housing is their shelter, their largest household consumption, and the main investment vehicle for them to accumulate wealth in the host cities. Spatial assimilation theory predicts that as immigrants are more socially integrated, their housing patterns increasingly resemble those of the native population (Turner & Hedman, 2014; Wessel et al., 2017). Empirical studies on spatial assimilation are mainly situated in economies with existing ethnic enclaves and assume spatial assimilation and residential segregation are "two sides of the same coin": socially integrated migrants will move from ethnic enclaves into natives' neighbourhoods, which is considered as a key milestone in spatial assimilation (Zorlu & Mulder, 2008; Andersen, 2010; Turner & Hedman, 2014).

However, there exist two important knowledge gaps regarding social integration and immigrants' housing behaviours. First, few studies disentangle the impact of social integration from ethnic segregation in the process of immigrants' spatial assimilation. Therefore, we cannot infer how social integration will impact immigrants' housing behaviours if host cities have few ethnic enclaves. The main reason for this gap is the absence of an institutional context capable of distinguishing confounding factors, such as residential segregation by ethnicity or nationality, to ascertain the causal impact of social integration on immigrants' housing behaviours.

Second, past studies usually focus on the snapshot of immigrants' housing behaviours when they arrive at host societies (Ihlanfeldt & Mayock, 2012; Ling et al., 2018), and few of them link social integration with the long-term housing behaviours from a temporal perspective, possibly due to a data gap: limited datasets have been able to concurrently record immigrants' social integration processes and their housing outcomes over a long time span (Coulter et al., 2016; Tanis, 2018). Therefore, it remains unclear whether social integration has a long-term impact on immigrants' housing behaviours beyond the first few years of settlement (Nowotny & Pennerstorfer, 2019; Zheng et al., 2020; Adam et al., 2021). More specifically, how immigrants' housing behaviours gradually evolve throughout the extended process of social integration, and whether they eventually converge with the natives' housing behaviours once the immigrants are well integrated, remain underexplored.

We address these gaps in the literature by exploring two research questions. First, we examine whether social integration impacts *yet-to-integrate* immigrants' housing behaviours over the post-migration years in cities where their housing choices are not significantly constrained by ethnic enclaves. Second, we investigate whether *well-integrated* immigrants exhibit housing patterns similar to those of natives, as predicted by the spatial assimilation theory. We focus on two key outcomes of housing behaviours in these research questions, including the commuting distances and the housing costs (prices or rent).

To investigate these research questions, we conducted a survey of 1,128 households living in Singapore's public housing market between 2017 and 2018, with the support of a local town council. The survey collected comprehensive data on immigrants' migration years, as well as both subjective and objective measures of social integration. The institutional features of Singapore's public housing market are particularly valuable for this study because of two reasons. First, the high ownership rate of public housing, and the strict public housing policies such as ethnic quotas at the building level, effectively prevent residential segregation by ethnicity in Singapore. It allows us to isolate the impact of social integration by minimizing confounding factors like the emergence of the ethnic enclaves, which is challenging in past studies. Second, comprehensive public housing rental and transaction data are available in Singapore, enabling us to match our self-collected survey data of immigrants' social integration with their rental and home-purchasing records. This helps bridge the data gap in previous literature and enables the investigation of the long-term impact of social integration on immigrants' housing behaviours over the post-migration years.

# 2. Literature Review and Institutional Background

## *2.1 Literature Review*

Immigrants' housing behaviours in host societies, including their residential location choices and the price or rent premiums they pay, have attracted extensive interest in the literature. Regarding residential location choices, Daniel McFadden—the 2000 Nobel Prize winner—introduced a classical model stating that residents weigh the attributes of each available alternative (e.g., accessibility to amenities, neighbourhood quality, cost of living, and dwelling characteristics) and choose the option that maximizes utility (McFadden, 1977). Since then, various socioeconomic factors, policy environments, and social networks have been documented to influence immigrants' residential choices. For instance, Dotti Sani & Acciai (2015) find that access to high-quality education for children is a key factor for immigrant families in selecting neighbourhoods in Los Angeles. Zapata-Barrero et al. (2022) show that immigrants are sensitive to local political attitudes and prefer neighbourhoods with more welcoming sentiments toward migration. Social networks also play a crucial role, as proximity to kin and migrant communities can provide essential support. Wang & Fan (2024) find that refugee migrants often settle in areas with existing refugee communities despite high unemployment rates.

Regarding immigrants' price/rent premiums, past studies mainly focus on the time when immigrants arrive in host societies and have documented several channels that drive these premiums. Specifically, the hedonic pricing model by Rosen (1974) assumes that the fair market price (or rent) of a house is the sum of the values of its characteristics (e.g., physical features and accessibility to services), while the differences between the actual transaction price (or rent) and the fair market price (or rent) are explained by idiosyncratic factors, such as immigrant buyers' information barriers. Combining the hedonic pricing model with a search model, Ihlanfeldt & Mayock (2012) show that nonlocal homebuyers pay 2% higher prices in Florida due to weaker bargaining power and less knowledge about the local market's nuances. Ling et al. (2018) argue that limited access to local intermediaries (brokers) explains a 3–10% price premium paid by immigrant homebuyers in major US metropolitan areas. Fan et al. (2023) find that immigrant homebuyers from mainland China pay higher prices in Hong Kong's housing market, with one explanation being weak bargaining power due to information barriers.

Our study focuses on the impact of social integration on immigrants' housing behaviours, including residential locations and price/rent premiums. Literature defines social integration as the process during which newcomers or minorities are incorporated into the social structure of the host society (Zheng et al., 2020). Developed as an extension of Gordon's (1964) social assimilation theory[1], spatial assimilation is considered one form of social assimilation and is achieved when immigrants adopt housing patterns similar to those of the native population (Zorlu & Mulder, 2008; Turner & Wessel, 2013). This phenomenon can be attributed to enhanced social integration such as improved social networks, better job opportunities, and aspirations for upward mobility (Warners, 1986).

Most empirical studies on spatial assimilation focus on economies with existing ethnic enclaves (Andersen, 2010; Shubin & Dickey, 2013) and examine two key outcomes: 1) where immigrants locate upon arrival, and 2) when immigrants make decisions to move out of ethnic enclaves. These studies show that new immigrants first choose to live in areas with dense immigrant populations (Zorlu & Mulder, 2008; Simone & Newbold, 2014). A sense of security in familiar cultural environments (Andersen, 2010) and informal networks play a key role in these initial residential location choices (Nowotny & Pennerstorfer, 2017). Over the post-migration years, immigrants gradually achieve spatial assimilation by moving closer to native residential patterns and leaving ethnic enclaves. Zorlu & Mulder (2008) document that immigrants in paid employment tend to relocate more frequently and move to less segregated neighbourhoods (Zorlu & Mulder, 2008). Leclerc et al. (2022) show that in the Netherlands, high social integration precedes citizenship acquisition, reducing housing market discrimination and increasing the likelihood of leaving ethnic enclaves. Nevertheless, these studies treat leaving ethnic enclaves as the milestone of completing spatial assimilation. It remains unclear how immigrants' housing behaviours evolve after leaving ethnic enclaves, and whether they eventually converge with the natives' housing behaviours in the long term.

[1] Social assimilation and social integration are two similar concepts. The differences lie in the fact that after social assimilation, immigrants no longer retain their own cultural identity, while after social integration, they still maintain their cultural identity (Gordon, 1964). In this article, since we do not focus on the extent to which immigrants retain their original culture, the two concepts are used interchangeably.

In this study, we aim to close two literature gaps. First, we extend the empirical studies on the outcomes of immigrants' spatial assimilation from cities with ethnic enclaves to those with little ethnic segregation. Different from prior studies using immigrants' relocation out of ethnic enclaves as an indicator of assimilation, we propose new outcomes for the spatial assimilation of labour immigrants in cities with few ethnic enclaves, which are commuting distance and rent/price premium. By taking advantage of Singapore's unique institutional settings, our paper extends the spatial assimilation theory to global cities with few foreign or ethnic enclaves.

Second, there is limited understanding of the longer-term impact of social integration on housing behaviours. Previous studies provide only a snapshot of the social integration at the arrival of migrants or during early years of settlement (Liu et al. 2023; Nowotny & Pennerstorfer, 2019). Also, there exists a data gap in simultaneously tracking immigrants' social integration processes and their long-term housing outcomes over time (Coulter et al., 2016). Using combined self-collected survey data with housing transaction records and new measures of social integration across different stages of integration, we are able to examine the long-term effect of social integration on housing behaviours of yet-to-integrate and well-integrated immigrants, respectively.

## *2.2. Institutional Background*

As a global city, Singapore welcomes large-scale influx of labour immigrants, to address labour shortages stemming from low fertility rates and economic expansion. In 2019, the ratio of total foreign labour force to resident labour force (citizens and permanent residents (PRs)) was 1 to 1.6. This institutional setting differs from migration studies in many other economies, where heterogeneous types of immigrants apart from labourers (e.g., humanitarian immigrants) co-exist (Zorlu & Mulder, 2008).

For decades, Singapore's immigration policy has classified these labour immigrants into two clearly distinguishable categories: foreign talents and foreign workers.[2] Foreign talents refer to foreigners with advanced skills in higher-paid professions (e.g., finance, IT, engineering, or artistic sectors) that add value to Singapore's economy and cultural capital. In the past, skilled professionals typically came from advanced Western countries. Nowadays, this group also includes individuals from China, India and other Southeast Asian countries, and the category of foreign talent has been expanded to include mid- and lower-level white collar positions. In total, the foreign talent immigrants account for 53% of Singapore's total foreign labour force.[3]

The rest of the foreign labour force comprises (low-skilled) foreign workers. Nearly half of them serve as a source of cheap labour (the data is extracted from the official statistics by Singapore Ministry of Manpower.), including domestic maids and manual workers in construction, marine shipyard and process sectors. Foreign workers normally live in dormitories provided by employers and have limited impact on local housing markets. Notably, immigration controls allow foreign workers only a limited period to stay in Singapore after their work contracts end, and their primary objective for coming to Singapore is to remit income to their families overseas. Therefore, our integration study focuses exclusively on foreign talents.

In Singapore, becoming a citizen signals the highest level of integration or a stage of full integration (this is also the case in the Netherlands as indicated in Leclerc et al. (2022)). Legally, the Singapore immigration authority takes two factors into account in assessing immigrants' applications for PR[4], and subsequently, citizenship: (1) the length of residency, (2) the level of integration into the society. To evaluate the integration level, the government assesses applicants' abilities to contribute to the Singapore economy (economic integration) and commitment to putting down roots (social integration). Before obtaining PR, an immigrant is likely to be in a "yet-to-integrate" stage. To assist yet-to-integrate immigrants to achieve the "milestones" of getting PR, the Singapore government has consistently taken initiatives to help immigrants integrate into society through designated government events and participation in organised community-based integration activities.

[2] Different visas are granted for foreign talents (employment passes and skilled passes) and foreign workers (work permits).
[3] The data is extracted from the official statistics by Singapore Ministry of Manpower.
[4] In Singapore, obtaining PR is the required intermediate step before immigrants can apply for citizenship.

One incentive for foreign talents to transfer to PR or citizen status is the privileged access to public goods. Taking public housing as an example: in 2019, 81% of Singapore's PR/citizen lived in high quality public housing with a 90% homeownership rate (HDB, 2019).[5] The heavily subsidised new public housing requires the buyer(s) to be a citizen or a family with at least one citizen. Resale public housing, which comes with fair subsidies, requires the buyer(s) to be a citizen, a family with at least one citizen, or a family of all PRs.[6] The generous subsidies for public housing ownership result in few citizens/PRs in the public rental housing market, which is mainly occupied by yet-to-integrate immigrants. This differential housing policy allows us to identify yet-to-integrate immigrants in the public open rental housing market and well-integrated immigrants (PRs or naturalized citizens) in the public owner-occupied housing market.

Additionally, tenancy agreements are typically renewed annually at the prevailing market rate and landlords must ensure that tenants have valid immigration status. Otherwise, an immigrant tenant will lose their right to continue renting a flat and be evicted immediately. Because of these institutional obstacles in the rental housing market, immigrants are motivated to become PRs and subsequently citizens to purchase subsidised public housing, as the expensive private housing is out of reach for the majority. This incentivizes them to proactively integrate into Singaporean society in order to gain citizenship.

[5] The rest, in upper income groups, lived in the expensive private housing.

[6] A single PR cannot purchase public housing, either in new sales or resales. Apart from PR/citizenship, there are some other requirements for purchasing public housing. For example, single citizens can only purchase 2-room flats after they are over 35 years old.

There are no evident ethnic or immigrant enclaves in public rental housing market in Singapore, due to stringent government regulations. To prevent the formation of foreigner enclaves in public housing estates, a Non-Citizen Quota (NCQ) for renting out an entire public flat was implemented in 2014. The maximum proportion of flats in each public housing building or neighbourhood that can be rented to non-citizens is capped at 11% or 8% respectively (HDB, 2018). This is in addition to the Ethnic Integration Quota (EIQ) implemented in 1989, under which the distribution of homeowners' ethnicities in a building, neighbourhood and estate must roughly mirror the national averages. These policies ensure that the shares of migrants and locals, as well as residents across different ethnic groups, roughly mirror the national distribution at the building level, effectively preventing the formation of immigrant and ethnic enclaves in public housing estates. Therefore, this institutional setting provides us with a quasi-experiment to identify the impacts of social integration on the housing trajectory of immigrants beyond confounding factors such as residential segregation by ethnicity or nationality.

As a city-state, Singapore has a polycentric spatial structure, with the Central Business District (CBD) at the southernmost part of the city serving as the primary economic hub (see Appendix Figure 1), and several decentralized regional centres developed in other directions, such as Jurong East in the west, Tampines in the east, and Woodlands in the north. Suburbanization has been led by the HDB over the past few decades, forming self-sufficient residential towns equipped with schools, healthcare, and commercial hubs. The Mass Rapid Transit (MRT) network connects these residential towns to the CBD and other employment clusters. This planned suburbanization ensures that all neighbourhoods in Singapore benefit from an efficient transportation network and have access to high-quality amenities.

## 3. Hypothesis Development

We focus on two outcomes of immigrants' spatial assimilation—commuting distance and price (rent) premium—in Singapore's context with no evident ethnic or foreign enclaves. Immigrants' integration outcomes are hypothesized to differ at the yet-to-integrate stage and well-integrated stage. At the yet-to-integrate stage, we examine how the level of immigrants' social integration gradually impacts their commuting distances and housing rents in the public open rental housing market. At the well-integrated stage, we examine whether there are any differences in commuting distances and housing prices between locally born and naturalised citizens in public owner-occupied resale housing market.

For commuting distance at the yet-to-integrate stage, we form the following hypothesis:

*H1. Newly arrived (less integrated) labour immigrants in Singapore live closer to their workplace; and the better integrated they are, the longer their commuting distance to work will be.*

This hypothesis is developed from the spatial assimilation theory, which is adapted from the broader residential location choice theory. Specifically, the theory of residential location choice posits that a household's housing location choice is a trade-off between proximity to job location and amenities (Straszhem, 1987). The theory of spatial assimilation suggests that with better social integration, immigrants' housing patterns increasingly resemble those of the natives (Turner & Wessel, 2013). Guided by these theories, we expect new labour immigrants, who are unfamiliar with the host city, to place greater emphasis on their job locations when making home location choices upon arrival, leading them to live closer to their workplaces (Tanis, 2018).[7] With greater social integration, immigrants are likely to place more importance on being near other amenities, similar to natives, resulting in an increase in their commuting distances.

[7] Similar finding is documented in Tanis (2018) that newly arrived labour immigrant's first home location in Germany is close to the workplace. However, their subsequent residential locations are mainly the ethnic enclaves in Germany. Our setting in Singapore is different as Singapore has few ethnic enclaves, which allow us to identify whether labour immigrants' subsequent residential locations are influenced by their social integration (beyond the impact of residential segregation).

Considering the second outcome—housing price (rent) premium—at the yet-to-integrate stage, we have the following hypothesis:

*H2. Immigrants pay lower rents in the public rental housing market as they become more socially integrated, holding the housing features constant.*

This hypothesis is developed from past studies on information asymmetry in an inefficient housing market. As reviewed in Section 2, literature consistently shows that newly arrived immigrants are unfamiliar with the housing market and face information barriers, so they pay higher rents or housing prices than locals, ceteris paribus (Ihlanfeldt & Mayock, 2012; Ling et al., 2018). However, immigrants are expected to gradually accumulate housing market information during the social integration process (Warnes, 1986), which mitigates the initial information deficit. For instance, although newly arrived immigrants mainly contact property agents of the same ethnicity or cultural background (White & Hurdley, 2003), they may gradually become acquainted with local agents in the process of integrating. Ryan (2011) documents that in the post-migration years, immigrants build a social network which can provide them better access to market information. In addition, more social capital can be generated from social interactions in the residential community and workplace after immigrants have settled in the receiving society (Glaeser et al., 2002; Weller & Bruegel, 2009; Reynolds, 2013). Indeed, it is found that homebuyers with more social capital enjoy a price discount of 3–7% (Tu et al., 2017). Therefore, we hypothesise that in the Singapore context, the better the immigrants integrate into the society, the lower the rent premiums they pay.

At the well-integrated stage, we hypothesize that the well-integrated immigrants have similar housing outcomes to the natives. Fully integrated immigrants are expected to have the same commuting patterns as locals because they have achieved spatial assimilation. Also, they are unlikely to face additional information barriers compared to locals and are not expected to pay price premiums in the public resale housing market. As discussed in Section 2.2, we consider naturalised citizens to be fully integrated immigrants, because it requires a long-term progress of social integration with the Singaporean society before immigrants can get citizenship (Sinning, 2010; Leclerc et al., 2022). Therefore, our hypotheses for the two housing outcomes at the well-integrated stage are stated as follows:

*H3. Naturalised citizens (i.e., well-integrated migrants) have similar commuting distances relative to the natives (locally born citizens).*

*H4. Naturalised citizens do not pay price premiums relative to the natives in the public resale housing market.*

# 4. Methodology, Variables, and Data

## *4.1. Measures of Immigrants' Integration*

Past literature suggests multiple measures for assessing immigrants' social integration. In addition to the length of residency, which plays an important role in immigrants' housing careers (Abramsson et al., 2002), immigrants' integration into a host society can also be assessed based on their subjective sense of belonging and the extent of their participation in various social relationships (Gao & Lim, 2023; Zheng et al., 2020). Therefore, integration is a multidimensional concept encompassing residency period, sense of belonging, and interactions with local society (Holt-Lunstad & Lefler, 2020).

Accordingly, we develop three measures for assessing immigrants' integration levels in this study. The first is the length of an immigrant's residency in Singapore. The second is the frequency with which an immigrant has participated in government-organised community activities, which acts as an objective measure of their level of integration. In our household survey, we asked respondents to use a 5-point scale to rate their frequency of participation in the following types of community events: (i) voluntary services,

(ii) festival celebrations, and (iii) government policy dialogues, which are the main community activities funded by Singapore People's Association[8] and organised by the town councils included in our survey. We then calculated the summation of the standardised scores in each question, which serves as our second measure. Our third measure is an aggregate score derived from a separate survey question assessing an immigrant's subjective sense of integration in Singapore. Following the strategy proposed by Snel et al. (2006) for measuring social integration, we asked respondents to rate their sense of belonging to Singapore on a 5-point scale: 1 ("very weak"), 2 ("slightly weak"), 3 ("neutral"), 4 ("slightly strong"), and 5 ("strong"). The raw scores were then standardised to have a mean of zero and a standard deviation of one, serving as our measure of subjective integration.

The conceptual diagram in Figure 1 illustrates the relationships among our three measures of social integration, which capture distinct (though partly overlapping) dimensions. Residency length reflects opportunities to integrate through longer exposure; community event attendance reflects active integration efforts; and subjective belonging reflects individual perceptions. These measures may evolve in parallel but remain distinct. For example, migrants may integrate passively by residing longer without participating in community events, whereas those who frequently attend events may well understand the society without developing strong belonging. By incorporating all three dimensions, our analyses capture a fuller picture of how integration shapes housing behaviours, as measured by commuting distance and housing rent.

### *4.2. Regression Model*

The first hypothesis, referring to the impacts of integration level on an immigrant renter's commuting distance at the yet-to-integrate stage, is tested using Equation (1). We focus on immigrant renters because in Singapore's housing context, immigrants at the yet-to-integrate stage can only rent from the public open rental housing market (see Section 2.2).

[8] Singapore People's Association is a statutory board in Singapore that oversees social cohesion in the state.

$$log(Distance_{it}) = \beta_1 INT_{it} + X'_{it}\lambda + \mu_t + \rho_i + \epsilon_{it}, \quad (1)$$

where $log(Distance_{it})$ is the natural logarithm of the commuting distance for immigrant $i$ at time *t*. $INT_{it}$ indicates the level of integration for immigrant renter *i* at the time *t* of renting,[9] which is measured by the length of residency, objective or subjective integration levels. The estimate of $\beta_1$ is our coefficient of interest, which indicates the effect of integration on an immigrant's commuting distance and is expected to be positive and statistically significant. $X_{it}$ is a set of control variables, which include conventional demographic and socioeconomic variables such as age, ethnicity, gender, income, marital status, and binary variables indicating whether the immigrant has at least one child in Singapore or has ever changed residence. In addition, we include a set of dummy variables to control for the occupations of the immigrants, as commuting distances are likely to vary across different occupations. $\mu_t$ represents year-quarter fixed effects corresponding to the start time of rental contracts. $\rho_i$ represents the district fixed effect of district $i$, and $\epsilon_{it}$ is the error term. Standard errors are clustered at the building level.

The second hypothesis, stating the impacts of integration level on an immigrant's housing rent at the yet-to-integrate stage, is tested using Equation (2).

$$log(Rent_{it}) = \beta_1 INT_{it} + X'_{it}\lambda + U'_{it}\gamma + \mu_t + \rho_i + \epsilon_{it}, \quad (2)$$

where $log(Rent_{it})$ is the natural logarithm of the monthly rent that immigrant tenant $i$ pays in the leasing contract signed at time $t$. $INT_{it}$ and $X_{it}$ share the same definitions as in Equation (1): the level of integration for immigrant renter *i* at the time *t* of renting, and a set of demographic and socioeconomic control variables. $U_{it}$ is a set of hedonic variables, including floor, property age, building type, distance to central business districts, bus stops, subway stations, hospitals, and parks. $\mu_t$ and $\rho_i$ capture year-quarter and district fixed effects, respectively, as in Equation (1). Standard errors are also clustered at the building level. Integration enhances immigrants' local knowledge and increases their sources of information, which is expected to increase their ability to search for suitable housing and/or raise their bargaining power. Thus, the estimate of $\beta_1$ is expected to be negative and statistically significant.

[9] The rental contracts for public housing units are typically renewed annually with the maximum lease period of 2 years (HDB, 2018). Thus, all rental contracts in our sample were signed after 2015. Since the period between the survey time and the rental contract is short, we use an immigrant's frequency of attending community events and self-reported scores on integration at the survey time as the proxy for their integration level at the time of renting.

The third and fourth hypotheses state that at the well-integrated stage, the impact of integration level on immigrants' housing outcomes disappears, so there are no differences in commuting distances and housing prices between naturalised and native citizens. We test the third and fourth hypotheses using Equations (3) and (4), respectively:

$$log(Distance_{it}) = \beta_1 INT_{it} + \beta_2 INT_{it} \times Naturalised_{it} + \beta_3 Naturalised_{it} + X_i'\lambda + \mu_t + \rho_i + \epsilon_{it}. \quad (3)$$

$$log(Price_{it}) = \beta_1 INT_{it} + \beta_2 INT_{it} \times Naturalised_{it} + \beta_3 Naturalised_{it} + X_i'\lambda + U_{it}'\gamma + \mu_t + \rho_i + \epsilon_{it}. \quad (4)$$

Specifically, $Naturalised_{it}$ is a dummy variable with 1 indicating a naturalised citizen and 0 indicating a locally born citizen. We exclude homeowners who are PRs or foreign immigrants in this analysis and only compare naturalised citizens with locally born citizens, because PRs and foreign immigrants cannot purchase public housing independently like naturalised and locally born citizens. Definitions of other variables are the same as in Equations (1) and (2). Standard errors are clustered at the building level. The estimates of $\beta_1, \beta_2, \beta_3$ in Equations (3) and (4) are expected to be statistically insignificant.

### *4.3. Addressing Endogeneity Concerns*

The OLS estimates may suffer from endogeneity due to measurement error, reverse causality, or omitted variables. For the first explanatory variable (residency length), measurement error is less a concern, because residency length is based on respondents' reported year of first arrival in Singapore, which is less likely to have individual calculation or memory errors. Reverse causality is also unlikely, since variations in residency length in survey year $t$ depend only on arrival year $t_0$, which cannot be reversely influenced by current housing outcomes. The main concern is omitted variables, such as cultural preferences or job mobility, that affect integration and housing simultaneously. To mitigate this concern, we rely on a rich set of sociodemographic and occupational control variables, as well as the home-moving histories, in all our regression models.[10]

To further address the concern for omitted factors that may not be well captured by the control variables (e.g., personal preferences), in the models using residency length as the explanatory variable, we adopt an instrumental variable (IV) strategy. Specifically, we use the strictness of national immigration policy in the year before migrants' arrival as the IV for their residency length, and the strictness is proxied by the annual growth rate of Singapore's foreign workforce. To test the robustness of our results, we also use the foreign population's annual growth rate as an alternative IV in a separate regression. Since the instruments' variations are at the year level, we cluster the standard errors by year in the IV estimations.[11]

The IV is relevant because tighter immigration policies lower quotas and impose stricter requirements, delaying migrants' arrival and shortening their residency length at observed survey time. In Singapore, work visas can only be obtained through registered employers. When policies tighten, the government raises screening standards for both sponsors and applicants (Jumabhoy et al., 2016), increasing the likelihood of rejection. Applicants must either reapply later or find alternative sponsors, delaying arrival time and reducing observed residency length.

[10] We also consider impacts of the confounding factors at the city or district level, such as spatial polarisation and residential segregation. In Online Appendix C, we discuss how Singapore's public housing setting and the design of our empirical models allow us to rule out the major confounding factors.

[11] Because government data on the foreign workforce and foreign population is only available from 1992 and 1980, respectively, immigrants who arrived before these years are excluded from the corresponding IV regressions. As a result, the number of observations in the IV estimations is slightly smaller than in the OLS estimations.

We consider these variations in arrival time due to policy strictness are exogenous, since changes in immigration policy are top-down national decisions, which are unlikely to correlate with unobserved individual preferences. Similar IV strategies have been used in prior studies, such as compulsory schooling laws serving as instruments for individual education years (Angrist & Krueger, 1991).

Our IV satisfies exclusion restriction, as historical immigration policy strictness likely affects future housing outcomes only through its impact on residency length. A potential violation arises if historical immigration policies influence future housing market by altering housing supply as well, which may affect rents or prices independently of integration. We address this concern by including district × contract year-month fixed effects in robustness checks, which capture local housing supply variations at the contract time.

Finally, for the other two explanatory variables (subjective integration scores and community event participation), the IV strategy is not applied because the immigration policy strictness is not conceptually associated with these two measures. Instead, we rely on our comprehensive control variables to capture possible confounding factors in these models. We fully acknowledge the potential endogeneity of the two measures arising from individual subjective decisions after arrival at Singapore, and we consider them as complement to our IV estimation results.

### *4.4. Data Collection*

To measure an immigrant's level of integration when their housing decision was made, we conducted a household survey between December 2017 and May 2018. We also collected information on the socioeconomic characteristics of renters and buyers, together with the hedonic features of their residence. We then combined the survey data with the transaction price or rent records from the Singapore Real Estate Exchange (SRX) housing transaction database. The survey proceeded in three steps. We first obtained approval from the Singapore North-West Community Development Council (CDC) to survey households living in the public housing estates under its management. The North-West CDC is one of five town councils in Singapore, with distance to the central business district (CBD) ranging from 4.7 to 15.9 km (Appendix Figure 1).[12] The Council approved our survey in three HDB towns including Clementi, Queenstown, and Bukit Panjang, which were chosen because they are mature residential districts with a sufficient supply of public housing units to conduct our survey.

Second, to ensure that an immigrant's level of integration collected in the survey is a reliable proxy for their level of integration at the time that their housing decision was made, we chose the most recent transaction records of each unit in our study area from the SRX database. In Singapore, a typical rental contract lasts 1-2 years, and all rental contracts we chose started during the second quarter of 2015. This sample period also ensured that all immigrants made their rental choices under the same policy setting, as the government's latest update in their integration policy for the public open rental housing was enacted in 2014 (HDB, 2018). For public housing resale records, we restricted our sample to transactions that were completed within 5 years before the survey.[13] In total, this resulted in 4,198 addresses of recent rental transactions and 8,797 addresses of recent public housing resales in the three HDB towns we studied.

Third, surveyors[14] visited all the addresses we obtained on an evening of a weekday or on a weekend, and a survey was conducted only if one of the decision-makers in renting or buying the home was present and agreed to be interviewed. We set an upper limit for the number of responses collected in each building to ensure representativeness across buildings. We obtained 1,128 valid samples, which included 513 immigrant renters and 615 homeowners; the latter comprising naturalised citizens, locally born citizens,

[12] Details of the North-West CDC can be found at https://www.cdc.gov.sg/northwest.
[13] The government restricts resales of public housing units within 5 years, the 5-year time window ensures that the homeowners we surveyed are the same as the homebuyers in the matched transaction records as well as ensures we can obtain sufficient records because of low response rate.
[14] The surveyors were well-trained before going to the field to guarantee high-quality and homogenous interviews.

PRs who pair with other PRs/citizens, and foreign immigrants who pair with citizens (Table 1). The surveyed housing units were all located in typical medium- to high-rise public housing buildings in Singapore, with locations shown in Appendix Figure 2. Singapore's public housing is relatively homogenous in terms of configuration and quality (Fan et al., 2021), so the housing units we surveyed were representative of the physical environment of Singapore's public housing.

Regarding the measures of integration, our survey sample shows that immigrant renters have lived in Singapore for an average of 7.8 years, and naturalised citizens/PRs for an average of 14.9 years, consistent with the rationale that immigrants need to live for a sufficiently long period to integrate into the society before obtaining PR/citizenship (Ho & Tyson, 2011). The standardised score for subjective sense of belonging and objective frequency of attending community events is also higher for homeowners than renters. Immigrants in the three mature HDB towns we surveyed can represent the overall integration levels of immigrants in Singapore, because immigrants in Singapore do not sort into different neighbourhoods based on their integration levels, as discussed in Section 2.2. Both government reports (Mathews et al., 2019) and anecdotal evidence[15] validate that the distributions of immigrants' integration levels are similar across HDB towns.

Our survey also collected household information such as commuting distance, age, ethnicity, gender, education, income, occupation, marital status, and a binary variable indicating whether the immigrant had ever changed residence. Around 76.80% of the immigrant renters have monthly income between 4,000 and 8,000 SGD, and 83.82% of them hold a bachelor's degree or above. Around 19% of the immigrant renters work in the IT industry, and 13% of them provide professional, scientific, and technical services. These findings are consistent with our review in Section 2.2 that immigrant residents in public rental housing markets are professional or high-skilled workers. For immigrants with a length of stay of less than

[15] See: https://www.straitstimes.com/opinion/st-editorial/public-housing-a-social-glue-for-integration

10 years, around 68% of them are married and 65% of them have children. For immigrants with length of stay of more than 10 years, 77% of them are married and 75% of them have children. In summary, our sample is representative in basic demographic information, which is generally consistent with the national distribution (DOS, 2018). As explained in Section 4.2, due to the differences in marital status and family composition, we explicitly control for such demographic features captured by two dummy variables of being married and having children, in addition to age and gender. By doing so, we remove the confounding effect from demographic factors in earlier versus later stages of integration.

The matched SRX housing transaction data shows that in our sample, the average rent is 1,439 SGD (1,060 USD) per month and around 85% of sampled tenants rent an entire flat rather than an individual room. The market trends in rent and resale price in our sample, estimated using the hedonic model (Rosen, 1974), are also closely aligned with the national level indices over the same periods (Appendix Figure 3), verifying that our sample is representative of the nationwide public rental/resale housing market performance in Singapore. The average floor level is 8.2, and the average unit size is 97 square meters. Geolocation information on the buildings was extracted from "OpenStreetMap", and distance to facilities such as bus stops, subway stations, CBD, general hospitals, and parks was calculated using GIS tools. All variables are defined in Appendix Table 1.

# 5. Empirical Results

## *5.1. Yet-to-integrate Stage: Social Integration and Home Renter's Commuting Distance*

Table 2 reports OLS estimation results for the impact of immigrant renters' level of integration on their commuting distance following Equation (1), which tests our Hypothesis 1. In Columns (1) to (2), we report the OLS estimation results using years of post-migration residency as the explanatory variable. To address the concern for potential multicollinearity issues among control variables, we first exclude the sociodemographic and occupational controls in Column (1) and then report the results with full controls in Column (2).[16] We find that the estimated coefficients are similar in both magnitude and statistical significance level, which validates the reliability of our results. After including the full controls, we find that one additional post-migration year leads to a 1.20% increase in commuting distance.[17] The estimate is statistically significant at the 5% level.

We proceed with the IV estimations to further alleviate the endogeneity concern. Appendix Table 6 reports the first-stage results of the IV estimations. As expected, we find that looser migration policies, proxied by higher growth rates of the foreign workforce or foreign population, are associated with longer residential length, confirming the relevance condition. With a one–percentage point increase in the annual growth rate of the foreign workforce or foreign population, residential length increases by 0.42 years or 0.56 years, respectively. The estimates are statistically significant at the 1% level, and the F-statistics exceed 10, alleviating concerns about weak instruments.

With the strong first-stage results, we present second-stage IV estimates in Columns (3)-(4) of Table 2. Column (3) reports the result using the main IV of foreign workforce growth rate. We find that one additional year of post-migration residency increases commuting distance by 3.04%. The IV estimate is statistically significant at the 5% level. Taking the typical time before citizenship conversion as approximately 10 years (Ho & Tyson, 2011), our findings translate into a 30% increase in commuting distance between home and workplace before fully integrating. Column (4) reports the result of using the alternative IV of foreign population growth rate. Under this IV, 10 years of post-migration residence leads to a 28% increase in the commuting distance, and the estimate is statistically significant at the 5% level. Among the demographic and socioeconomic control variables, we find that higher incomes are associated with longer commuting distances, while we do not find statistically significant impact for factors like age, marital status, or whether they have children.

[16] To further alleviate concerns about multicollinearity, we conduct Variance Inflation Factor (VIF) tests for our regression models in Tables 2–5, with the corresponding results reported in Appendix Tables 2–5. We find that the VIFs for all independent and control variables in our regression models are below 10, the conventional threshold for detecting significant multicollinearity issues. This set of evidence mitigates the concerns on multicollinearity among independent variables.

[17] We confirm that outcome variables in all our regression models do not contain zero values, so the estimates can be reliably interpreted as the percentage change (Chen & Roth, 2022).

In Appendix Tables 7-8, we show that our OLS and IV results are robust to the inclusion of district × contract year-month fixed effects, which can capture the district-level variations in housing supply. Further, with fewer clusters, standard errors estimated via conventional clustering methods are more likely downward-biased. We address this concern by using the Wald bootstrap inference to re-calculate the statistical power, as presented in Appendix Tables 9-10. Our estimates remain robust.

In Columns (5)-(6) of Table 2, we use the standardised frequency of attending community events as the independent variable, which reflects the immigrant's objective level of integration. Sociodemographic and occupational control variables are excluded in Column (5) and are included in Column (6). We find that if an immigrant renter's frequency of attending community events increases by one standard deviation, their commuting distance increases by 7.60% (Column (5)) to 8.81% (Column (6)). The estimates are statistically significant at the 5% level.

In Columns (7)-(8), the explanatory variable is the standardised score for the immigrant's subjective sense of belonging to Singapore. We continue to observe positive coefficients in models either without or with the control variables, although the magnitudes are small relative to the standard errors and are statistically insignificant. One possible reason for the weak statistical power is that the subjective sense of belonging is measured based on the survey responses to only one single question on a 5-point scale, which provides less detailed variations compared to our objective measure that combines multiple questions of participation frequencies in different community events. Additionally, compared to the objective measures, subjective reporting of the sense of belonging may have relatively larger measurement errors, potentially due to personal biases or interpretation differences. Both reasons could result in inflated standard errors and statistically insignificant estimates. We leave this for future investigation, when more advanced data is available.

In summary, this empirical evidence indicates that immigrant renters have longer commuting distances if they are better integrated into the host society. Specifically, new immigrants tend to look for housing near their workplace upon arrival. However, with time, they develop a stronger sense of integration with Singaporean society, and attend local community activities more frequently. They then start showing more heterogeneity in housing demand, and search for other housing opportunities farther from their workplace. More distant housing may offer benefits such as better community connections, better schooling for their children, or more affordable rent.

### *5.2. Yet-to-integrate Stage: Social Integration and Housing Rent*

Table 3 presents the estimated impacts of immigrant renters' integration on housing rent using Equation (2), which tests our Hypothesis 2.[18] The OLS estimation results without and with control variables are reported in Columns (1)-(2), respectively, which are consistent. After controlling for sociodemographic features of renters and physical features of the unit, we find that with 10 more years of residency after immigration, immigrant tenants pay 2.2% less rent. The estimate is statistically significant at the 5% level. Columns (3) and (4) report estimation results using the growth rate of the foreign workforce and foreign population as an IV, separately. Given 10 additional years of post-migration residency, immigrant tenants are estimated to pay 6.7% and 5.6% lower rent, respectively. Both IV estimates are statistically significant at the 1% level.

[18] We exclude samples who sublet one room in a flat due to missing information on room size.

In Columns (5) and (6), we use the standardised objective frequency of attending community events as the independent variables, while in Columns (7) and (8), the measure of subjective sense of belonging to Singapore is used. Control variables are excluded in Columns (5) and (7), and are included in Columns (6) and (8). Considering the estimates with full controls, we find if the objective and subjective levels of integration increase by one standard deviation, the rent paid by immigrants decreases by 1.08% and 1.23%, respectively, with statistical significance at conventional levels.[19] Results without control variables are consistent in both magnitudes and statistical significance level.

In summary, our results show that immigrants indeed pay lower rent after they reside in Singapore for a longer period. They also pay lower rent if they hold a stronger sense of integration or attend local community events more frequently. While past literature documents that newly arrived immigrants pay a higher housing price or rent than locals by 3–8% in various global housing markets (Ling et al., 2018; Tu et al., 2017), we provide new insights that these price disparities gradually diminish as immigrants become better integrated with the society over the post-migration years.

### *5.3. Well-integrated Stage: Social Integration and Homeowners' Commuting Distance*

We proceed to test Hypothesis 3 using Equation (3), with the corresponding results reported in Table 4. Specifically, we compare the commuting distances of naturalised citizens (i.e., well-integrated migrants) with those of locally born citizens (i.e., natives). Since the length of post-migration residency is not applicable to locally born citizens, we use only the frequency of attending community events and the subjective sense of belonging to Singapore as measures of integration in this analysis. To check the consistency of the estimates, we either exclude or include sociodemographic and occupational control variables in separate models. We find confirming evidence that differences in commuting distances between well-integrated migrants and natives are not statistically significant, as indicated by the statistically insignificant coefficients for the naturalised citizen dummy variable, the integration measures ("Attend Community Events" or "Subjective Integration"), and their interaction terms in all models.

[19] Our findings remain robust after considering transportation costs, saving effects, and unit size.

While well-integrated immigrants have commuting distances similar to those of natives, yet-to-integrate immigrants have much shorter commuting distances than both natives and well-integrated immigrants. The summary statistics in Table 1 show that home renters (i.e., yet-to-integrate immigrants) have an average commuting distance of 7.24 km, whereas homeowners (i.e., locally born natives and well-integrated immigrants who became naturalised citizens) have an average commuting distance of 9.44 km. After controlling for the sociodemographic characteristics, occupations, and district and time fixed effects, we estimate that the commuting distances of home renters are 22.2% shorter than those of homeowners, as reported in Appendix Table 11.

In summary, considering the regression results in Tables 2 and 4, we find that, conditional on same demographic factors (e.g., age), newly arrived immigrants commute shorter distances than the locals. Their commuting distances gradually increase with better integration until they become fully integrated, at which point their commuting distances become similar to those of natives with same sociodemographic characteristics.

### 5.4. *Well-integrated Stage: Social Integration and Housing Price*

Last, to test our Hypothesis 4, we compare the housing prices paid by naturalised and locally born citizens, using Equation (4). The results are reported in Table 5. We use a dummy variable to denote naturalised citizens and interact the dummy variable with the measurements of the homeowner's integration level—the standardised frequency of attending community events (Columns (1) and (2))—or their subjective sense of belonging to Singapore (Columns (3) and (4)). Control variables of sociodemographic and housing features are excluded in Columns (1) and (3) and are included in Columns (2) and (4). As expected, the estimated coefficients for the dummy variable of naturalised citizen, the measurements of integration, and the interaction terms are all statistically insignificant in all models. This empirical evidence shows that after immigrants are well-integrated into the host society and convert to citizens, there are no significant differences in the home purchasing prices they pay compared to those paid by locally born citizens, who have same sociodemographic characteristics (e.g., in the same age).

In summary, our results in Table 3 show that better-integrated immigrants pay lower rents, and Table 5 demonstrates that once they are well integrated and become eligible to own public resale housing, they no longer pay price premiums over natives.[20] Taken together, these findings support our hypothesis that better social integration enables immigrants to gradually overcome information barriers, thereby reducing the housing cost premiums they initially face. By adopting a temporal perspective, these new findings build upon previous literature, which documents that newly arrived immigrants pay higher prices than local homebuyers due to factors such as information disadvantages (Ihlanfeldt & Mayock, 2012; Ling et al., 2018).

## 6. Discussion and Conclusion

In this paper, we examine the temporal effect of immigrants' social integration on housing behaviours at the yet-to-integrate and well-integrated stages separately, using Singapore's differential public housing policies as a quasi-natural experiment. Combining self-collected survey data on immigrants' social integration with recent rental and purchase transaction records, our study has two main findings. First, we find that social integration has a temporal impact on immigrants' spatial assimilation, which lasts for years until they become well integrated. In cities with little ethnic residential segregation, like Singapore, the more an immigrant renter integrates into the host society, the farther they choose to live from their workplace, and the lower the housing rents they pay, ceteris paribus. Specifically, after 10 post-migration years—the typical time before converting to citizens (Ho & Tyson, 2011), immigrants experience a 30% increase in commuting distance and a 6.7% decrease in rents. Similar results are observed when using a sense of belonging to Singaporean society or the frequency of attending local community activities as alternative measures of social integration. Second, we find that at the well-integrated stage, immigrants demonstrate similar housing patterns to natives, with negligible differences in commuting distances and property prices. In Singapore's context, the sample of naturalised citizen homeowners—who have demonstrated full integration before they can obtain citizenship—shows no significant impact of social integration on their housing behaviours any further.

[20] Since the public rental housing market in Singapore primarily accommodates yet-to-integrate immigrants and there are very few locals rent public housing units, we cannot directly compare the rent paid by yet-to-integrate immigrants with the hypothetical rent that locals would pay if they were in the rental market. Similarly, because public housing units can only be purchased by well-integrated immigrants and natives, we cannot compare the prices paid by yet-to-integrate immigrants and natives as homeowners.

To better illustrate the policy relevance of our findings, we examine heterogeneity across immigration subgroups. Given that the majority of Singapore's citizens are Chinese, we split migrant tenants in public housing into Chinese (majority) and non-Chinese (minority) groups and re-estimate the models in Tables 2 and 3. Results in Appendix Tables 12-13 show that the temporal effects of integration on housing outcomes (commuting distance and rent) are stronger among non-Chinese immigrants, while effects for Chinese immigrants are statistically insignificant. This likely reflects lower initial integration barriers for Chinese migrants (e.g., language in navigating the housing market). Thus, our findings suggest that integration policies should pay particular attention to minority groups, as integration exerts a stronger influence on their long-term housing outcomes.

Our study is closely linked to the spatial assimilation theory. The theory predicts that as immigrants become socially integrated, they adopt housing patterns similar to those of the natives (Turner & Hedman, 2014; Wessel et al., 2017). Most empirical evidence of the spatial assimilation theory comes from the economies with clear ethnic residential segregation and uses immigrants' relocation from ethnic enclaves into natives' neighbourhoods as a key milestone (Zorlu & Mulder, 2008; Andersen, 2010). We extend the empirical studies of spatial assimilation theory by showing that in cities with little ethnic residential segregation, different levels of social integration result in divergent housing patterns among immigrants, as measured by new indicators of commuting distances to workplaces and housing rents/prices. Immigrants who have a lower level of social integration live closer to workplaces and pay higher rent premiums in the housing market, ceteris paribus. Nevertheless, immigrants' housing patterns will converge with those of the locals (i.e., achieving spatial assimilation) when they are fully integrated into the society. The institutional setting in Singapore enables us to rule out the impact of confounding factors in identifying the causal effect of social integration on immigrants' housing behaviours. In other words, we consider Singapore's distinct setting as a social laboratory, with otherwise confounding factors, such as ethnic segregation or political preferences, well under control.

While Singapore's institutional setting is unique, the research question—how social integration impacts labour immigrants' housing behaviours, especially in cities with no obvious ethnic segregation—is universal and important in many other global cities. For instance, there is massive labour-oriented internal migration among Chinese cities (Han & Li, 2017), with little ethnic segregation in host cities. The housing behaviours of these immigrants are important for local socioeconomic development (Wu, 2002). Although the magnitudes of social integration's effect on housing behaviours can vary across countries/regions, the positive effect of social integration on commuting distances and the negative effect on rents along with the integration process are applicable to other housing contexts *without* ethnic enclaves.

In addition, using citizenship as a milestone of full integration is not unique to Singapore. The same mindset is applicable in Western countries such as the Netherlands and the U.S. (Wyman, 1996; Leclerc et al., 2022). Thus, our method of using citizenship as a watershed to distinguish between yet-to-integrate and well-integrated stages and estimate the impact of social integration on housing behaviours separately can be extended to other housing contexts.

The integration of immigrants is crucial for social sustainability. Yet, most prior academic research and policy discourse focus on integration policies as a snapshot of immigrants' initial arrival or during early years of settlement in the host countries (Zheng et al., 2020; Adam et al., 2021). Our results show a new temporal dimension, which is often neglected in the literature but is important in estimating the longer-term impact of social integration on housing behaviours. We illustrate that differences in housing trajectories between immigrants and natives continue to evolve for years after immigration, until they reach the milestone of becoming naturalised citizens. Government integration programs, such as organizing community events, are effective instruments for facilitating migrants' housing integration in post-migration years. Therefore, we advocate for integration policies that extend over a longer term, rather than merely offering support upon arrival or during the initial stages of immigration. For example, beyond one-time orientation programs, local governments are encouraged to invest in longer-term community events, networking initiatives, and professional upskilling workshops to strengthen immigrants' sense of belonging. Additionally, since information barriers contribute to housing cost premiums for yet-to-integrate immigrants, policymakers could establish housing advisory services or digital platforms tailored to different stages of integration, providing market insights, mortgage guidance, and/or housing policy information. Last, our heterogeneity analysis results indicate that minority ethnic groups benefit most from social integration, pointing to the need for integration policies towards them.

# References

Abramitzky, R., & Boustan, L. (2017). Immigration in American Economic History. *Journal of Economic Literature, 55*(4), 1311-1345.

Abramsson, M., Borgegard, L.-E., & Fransson, U. (2002). Housing Careers: Immigrants in Local Swedish Housing Markets. *Housing Studies, 17*(3), 445-464.

Adam, F., Föbker, S., Imani, D., Pfaffenbach, C., Weiss, G., & Wiegandt, C. C. (2021). "Lost in transition"? Integration of refugees into the local housing market in Germany. *Journal of Urban Affairs*, *43*(6), 831-850.

Andersen, H. S. (2010). Spatial Assimilation in Denmark? Why do Immigrants Move to and from Multi-ethnic Neighbourhoods? *Housing Studies, 25*(3), 281-300.

Angrist, J. D., & Krueger, A. B. (1991). Does compulsory school attendance affect schooling and earnings? *Quarterly Journal of Economics*, 106(4), 979–1014.

Boeri, T., De Philippis, M., Patacchini, E., & Pellizzari, M. (2015). Immigration, Housing Discrimination and Employment. *The Economic Journal, 125*(586), F82-F114.

Candipan, J., Phillips, N. E., & Small, M. (2021). From residence to movement: The nature of racial segregation in everyday urban mobility. *Urban Studies, 58*(15), 3095-3117.

Chen, J., & Roth, J. (2024). Logs with zeros? Some problems and solutions. *The Quarterly Journal of Economics*, *139*(2), 891-936.

Coulter, R., van Ham, M., & Findlay, A. M. (2016). Re-thinking residential mobility: Linking lives through time and space. *Progress in Human Geography, 40*(3), 352-374.

Cutler, D. M., Glaeser, E. L., & Vigdor, J. L. (2008). When are ghettos bad? Lessons from immigrant segregation in the United States. *Journal of Urban Economics, 63*(3), 759-774.

McFadden, D. (1977). Modelling the choice of residential location. Cowles Foundation for Research in Economics, Yale University.

DOS. (2018). *Population Trends, 2018.* Singapore: Department of Statistics, Ministry of Trade & Industry.

Dotti Sani, G. M., & Acciai, C. (2015). The residential location choice of immigrant parents in the Los Angeles metropolitan area. *Journal of Population Economics*, 28, 911–936.

Fan, Y., Hu, M. R., Wan, W. X., & Wang, Z. (2023). A Tale of Two Cities: Mainland Chinese Buyers in the Hong Kong Housing Market. *Review of Finance*, rfad006.

Fan, Y., Teo, H. P., & Wan, W. X. (2021). Public transport, noise complaints, and housing: Evidence from sentiment analysis in Singapore. *Journal of Regional Science, 61*(3), 570-596.

Gao, T., & Lim, S. (2023). Socio-spatial integration in innovation districts: Singapore's mixed-use experiment. *Cities*, 140, 104405.

Glaeser, E. L., Laibson, D., & Sacerdote, B. (2002). An Economic Approach to Social Capital. *The Economic Journal, 112*(483), F437-F458.

Gordon, M. (1964). *Assimilation in American life: The role of race, religion, and national origins*. Oxford University Press, USA.

Han, J., & Li, S. (2017). Internal migration and external benefit: The impact of labor migration on the wage structure in urban China. *China Economic Review*, *46*, 67-86.

HDB. (2018). *Regulations for renting out your flat: Period.* Housing and Development Board Singapore.

HDB. (2019). *HDB 2018/2019 Annual Report: Key Statistics.* Housing and Development Board Singapore.

Ho, Y. J., & Tyson, A. D. (2011). Malaysian migration to Singapore: Pathways, mechanisms and status. *Malaysian Journal of Economic Studies, 48*(2), 131-145.

Holt-Lunstad, J., & Lefler, M. (2020). Social Integration. In D. Gu, & M. E. Dupre, *Encyclopedia of Gerontology and Population Aging* (pp. 1-11). Springer, Cham.

Ihlanfeldt, K., & Mayock, T. (2012). Information, Search, and House Prices: Revisited. *Journal of Real Estate Finance and Economics, 44*(1-2), 90-115.

Jumabhoy, F., Lumsdaine, T., & Yazid, S. (2016). Singapore: Tightening the rules on hiring foreign workers. Lexology. https://www.lexology.com/library/detail.aspx?g=fb445065-4094-453e-afee-ba45a25444cc

Leclerc, C., Vink, M., & Schmeets, H. (2022). Citizenship acquisition and spatial stratification: Analysing immigrant residential mobility in the Netherlands. *Urban Studies, 59*(7), 1406-1423.

Ling, D. C., Naranjo, A., & Petrova, M. T. (2018). Search Costs, Behavioral Biases, and Information Intermediary Effects. *Journal of Real Estate Finance and Economics, 57*(1), 114-151.

Liu, T., Shi, Q., & Zhuo, Y. (2023). Are adaptation challenges relevant to the location choices of internal migrants? Evidence from China. *Regional Studies*, 1-13.

Mathews, M., Tay, M., & Teo, K. (2019). *Integral - A Report on Social Integration in Singapore for the 10th Anniversary of the National Integration Council.* Singapore: Institute of Policy Studies.

Nicholls, W. J., Uitermark, J., & van Haperen, S. (2020). Going national: how the fight for immigrant rights became a national social movement. *Journal of Ethnic and Migration Studies, 46*(4), 705-727.

Nowotny, K., & Pennerstorfer, D. (2019). Network migration: do neighbouring regions matter? *Regional Studies*, 53(1), 107-117.

Reynolds, T. (2013). 'Them and Us': 'Black Neighbourhoods' as a Social Capital Resource among Black Youths Living in Inner-city London. *Urban Studies, 50*(3), 484-498.

Rosen, S. (1974). Hedonic Prices and Implicit Markets: Product Differentiation in Pure Competition. *Journal of Political Economy, 82*(1), 34-55.

Ryan, L. (2011). Migrants' Social Networks and Weak Ties: Accessing Resources and Constructing Relationships Post-Migration. *The Sociological Review, 59*(4), 707-724.

Sassen, S. (1991). *The Global City: New York, London, Tokyo.* Princeton, NJ: Princeton University Press.

Schmidt Sr., R. (2007). Comparing Federal Government Immigrant Settlement Policies in Canada and the United States. *American Review of Canadian Studies, 37*(1), 103-122.

Simone, D., & Newbold, B. (2014). Housing Trajectories Across the Urban Hierarchy: Analysis of the Longitudinal Survey of Immigrants to Canada, 2001–2005. *Housing Studies, 29*(8), 1096-1116.

Sinning, M. (2010). Homeownership and Economic Performance of Immigrants in Germany. *Urban Studies, 47*(2), 387-409.

Snel, E., Engbersen, G., & Leerkes, A. (2006). Transnational involvement and social integration. *Global Networks, 6*(3), 285-308.

Straszhem, M. (1987). The Theory of Urban Residential Location. *Handbook of Regional and Urban Economics, 2*(1), 717-757.

Tu, Y., Li, P., & Qiu, L. (2017). Housing search and housing choice in urban China. *Urban Studies, 54*(8), 1851-1866.

Tanis, K. (2018). Regional distribution and location choices of immigrants in Germany. *Regional Studies*, 54(4), 483–494.

Turner, L. M., & Hedman, L. (2014). Linking Integration and Housing Career: A Longitudinal Analysis of Immigrant Groups in Sweden. *Housing Studies, 29*(2), 270-290.

Turner, L. M., & Wessel, T. (2013). Upwards, Outwards and Westwards: Relocation of Ethnic Minority Groups in the Oslo Region. *Geografiska Annaler. Series B, Human Geography, 95*(1), 1-16.

van der Laan Bouma-Doff, W. (2007). Confined Contact: Residential Segregation and Ethnic Bridges in the Netherlands. *Urban Studies, 44*(5/6), 997-1017.

Wang, W., & Fan, C. C. (2024). Spatial overlap: trade-offs in refugees' residential choices. *Journal of Ethnic and Migration Studies*, 50(2), 345–363.

Weller, S., & Bruegel, I. (2009). Children's `Place' in the Development of Neighbourhood Social Capital. *Urban Studies, 46*(3), 629-643.

Wessel, T., Andersson, R., Kauppinen, T., & Andersen, H. S. (2017). Spatial integration of immigrants in Nordic cities: The relevance of spatial assimilation theory in a welfare state context. *Urban Affairs Review*, *53*(5), 812-842.

White, P., & Hurdley, L. (2003). International Migration and the Housing Market: Japanese Corporate Movers in London. *Urban Studies, 40*(4), 687-706.

Wu, W. (2002). Migrant housing in urban China: Choices and constraints. *Urban Affairs Review*, *38*(1), 90-119.

Wyman, M. (1996). In *Round Trip to America: The Immigrants Return to Europe, 1880-1930* (pp. 99-100). Ithaca, NY: Cornell University Press.

Zapata-Barrero, R., Caponio, T., & Scholten, P. (2022). How political reception contexts shape location decisions of immigrants: Evidence from Spain. *Journal of Ethnic and Migration Studies*, 48(18), 4401–4419.

Zheng, S., Song, Z., & Sun, W. (2020). Do affordable housing programs facilitate migrants' social integration in Chinese cities? *Cities*, *96*, 102449.

Zorlu, A., & Mulder, C. H. (2008). Initial and Subsequent Location Choices of Immigrants to the Netherlands. *Regional Studies, 42*(2), 245-264.

**Figure 1. Conceptual Diagram**

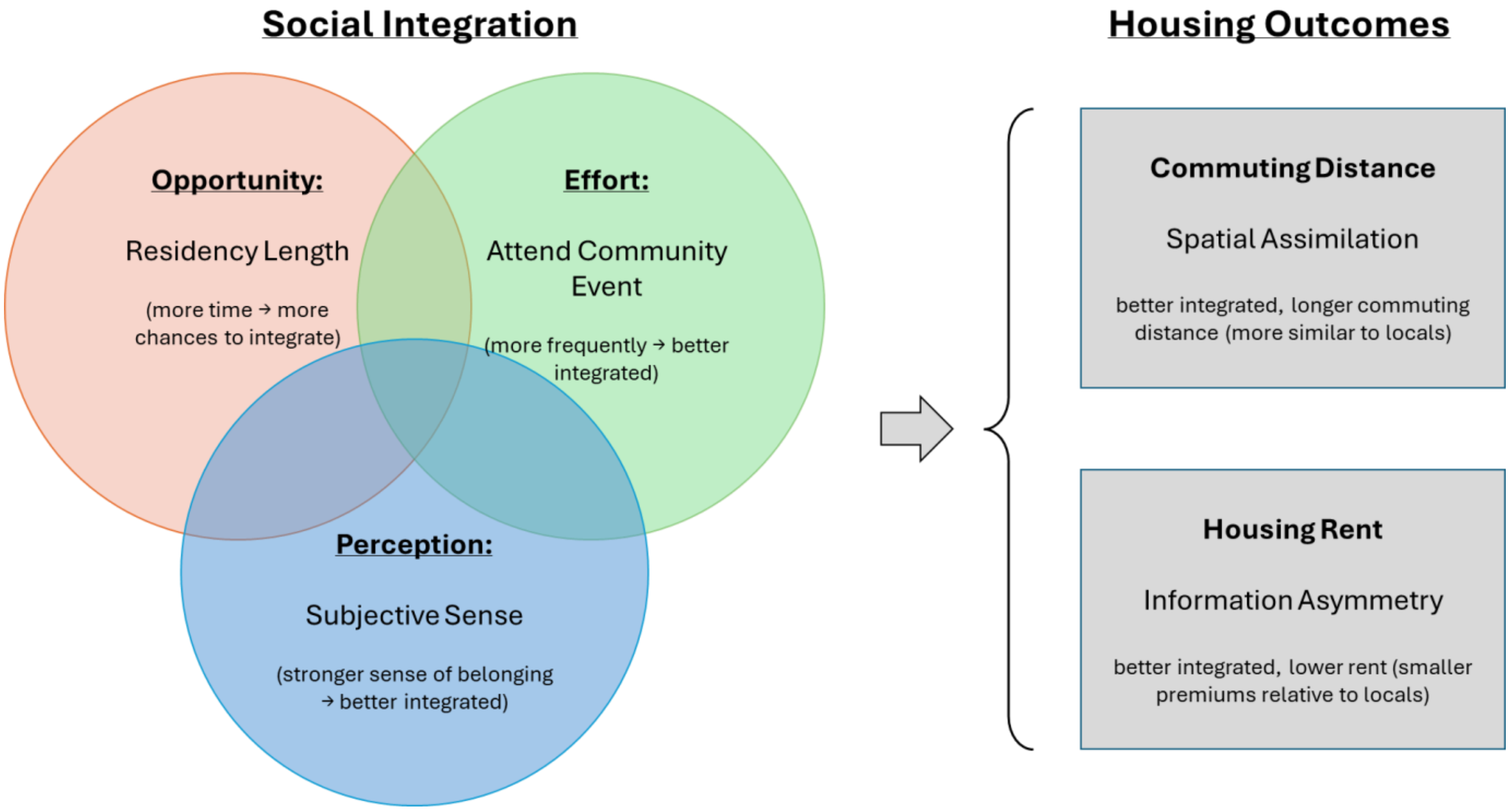

**Table 1: Summary Statistics**

**Panel A. Continuous and Indicatory Variables**

| | (1) | (2) | (3) | (4) | (5) | (6) | (7) | (8) | (9) |
|---|---|---|---|---|---|---|---|---|---|
| | | All Samples | | | Renters | | | Owners | |
| | Obs. | Mean | S.D. | Obs. | Mean | S.D. | Obs. | Mean | S.D. |
| **Personal Attributes** | | | | | | | | | |
| Age | 1,128 | 38.233 | 9.451 | 513 | 34.470 | 7.404 | 615 | 41.372 | 9.825 |
| Immigrant | 513 | 1 | 0 | 513 | 1 | 0 | | | |
| Naturalised Citizen | 615 | 0.138 | 0.345 | | | | 615 | 0.138 | 0.345 |
| Immigration Years | 790 | 10.268 | 7.944 | 513 | 7.788 | 6.642 | 277 | 14.863 | 8.119 |
| Immigration Years at Purchase Time | | | | | | | 277 | 10.170 | 8.076 |
| Subjective Integration | 1,128 | 0 | 1 | 513 | -0.123 | 1.023 | 615 | 0.103 | 0.969 |
| Attend Community Events | 1,128 | 0 | 1 | 513 | -0.052 | 0.961 | 615 | 0.043 | 1.030 |
| Gender (Male = 1) | 1,128 | 0.618 | 0.486 | 513 | 0.704 | 0.457 | 615 | 0.546 | 0.498 |
| Married | 1,128 | 0.767 | 0.423 | 513 | 0.700 | 0.459 | 615 | 0.823 | 0.382 |
| With Child | 1,128 | 0.747 | 0.434 | 513 | 0.789 | 0.408 | 615 | 0.712 | 0.453 |
| Ever Moved Home | 1,128 | 0.727 | 0.446 | 513 | 0.714 | 0.453 | 615 | 0.738 | 0.440 |
| **Housing Attributes** | | | | | | | | | |
| Rent (SGD) | 513 | 1,438.591 | 875.011 | 513 | 1,438.591 | 875.011 | | | |
| Rent a Room in a Flat | 513 | 0.150 | 0.358 | 513 | 0.150 | 0.358 | | | |
| Price (thousand SGD) | 615 | 436,117 | 102,214 | | | | 615 | 436,117 | 102,214 |
| Commuting Distance (km) | 1,128 | 8.438 | 5.379 | 513 | 7.239 | 5.156 | 615 | 9.438 | 5.361 |
| Unit Size (sq. m.) | 974 | 96.949 | 24.046 | 362 | 90.127 | 25.416 | 612 | 100.984 | 22.254 |
| Floor | 1,128 | 8.158 | 5.344 | 513 | 8.343 | 5.326 | 615 | 8.003 | 5.359 |
| Building Age | 1,124 | 26.465 | 11.359 | 511 | 32.323 | 11.656 | 613 | 21.582 | 8.427 |
| Distance to Bus Stop (km) | 1,128 | 0.119 | 0.057 | 513 | 0.122 | 0.058 | 615 | 116.199 | 55.084 |
| Distance to CBD (km) | 1,128 | 11.875 | 3.044 | 513 | 10.030 | 3.351 | 615 | 13.413 | 1.560 |
| Distance to Hospital (km) | 1,128 | 5.243 | 3.182 | 513 | 3.611 | 3.067 | 615 | 6.604 | 2.581 |
| Distance to MRT (km) | 1,128 | 0.348 | 0.198 | 513 | 0.431 | 0.211 | 615 | 0.278 | 0.154 |
| Distance to Park (km) | 1,128 | 0.258 | 0.154 | 513 | 0.297 | 0.180 | 615 | 0.225 | 0.118 |
| Executive Condominium | 1,128 | 0.081 | 0.272 | 513 | 0.037 | 0.189 | 615 | 0.117 | 0.322 |

**Panel B. Categorical Variables**

| | (1) | (2) | (3) | (4) | (5) | (6) |
|---|---|---|---|---|---|---|
| | All Samples | | Renters | | Owners | |
| | Count | Percent | Count | Percent | Count | Percent |
| ***Citizen status*** | | | | | | |
| Foreigner | 429 | 38.03% | 392 | 76.41% | 37 | 6.02% |
| Singapore PR | 276 | 24.47% | 121 | 23.59% | 155 | 25.20% |
| Naturalised Citizen | 85 | 7.54% | 0 | 0.00% | 85 | 13.82% |
| Local-born Citizen | 338 | 29.96% | 0 | 0.00% | 338 | 54.96% |
| ***Race*** | | | | | | |
| Chinese | 530 | 46.99% | 157 | 30.60% | 373 | 60.65% |
| Indian | 326 | 28.90% | 212 | 41.33% | 114 | 18.54% |
| Malay | 86 | 7.62% | 5 | 0.97% | 81 | 13.17% |
| Other | 186 | 16.49% | 139 | 27.10% | 47 | 7.64% |
| ***Education Level*** | | | | | | |
| Secondary school or below | 179 | 15.87% | 25 | 4.87% | 154 | 25.04% |
| Junior college or polytechnic | 229 | 20.30% | 58 | 11.31% | 171 | 27.80% |
| Bachelor’s degree | 475 | 42.11% | 268 | 52.24% | 207 | 33.66% |
| Postgraduate degree | 245 | 21.72% | 162 | 31.58% | 83 | 13.50% |
| ***Income*** | | | | | | |
| Below S$2,000 | 35 | 3.10% | 24 | 4.68% | 11 | 1.79% |
| S$2,001-S$4,000 | 133 | 11.79% | 41 | 7.99% | 92 | 14.96% |
| S$4,001-S$8,000 | 859 | 76.15% | 394 | 76.80% | 465 | 75.61% |
| Above S$8,000 | 101 | 8.95% | 54 | 10.53% | 47 | 7.64% |
| ***Occupation Industry*** | | | | | | |
| Accommodation and Food Services | 48 | 4.26% | 17 | 3.31% | 31 | 5.04% |
| Arts, Entertainment and Recreation | 17 | 1.51% | 5 | 0.97% | 12 | 1.95% |
| Education | 81 | 7.18% | 29 | 5.65% | 52 | 8.46% |
| Electricity, Gas, Steam and Air-Condition | 59 | 5.23% | 17 | 3.31% | 42 | 6.83% |
| Financial, Real Estate and Insurance | 81 | 7.18% | 38 | 7.41% | 43 | 6.99% |
| Health and Social Services | 90 | 7.98% | 49 | 9.55% | 41 | 6.67% |
| Information and communications | 131 | 11.61% | 97 | 18.91% | 34 | 5.53% |
| Manufacturing and Construction | 182 | 16.13% | 84 | 16.37% | 98 | 15.93% |
| Others | 204 | 18.09% | 85 | 16.57% | 119 | 19.35% |
| Professional, Scientific, and Technical | 105 | 9.31% | 66 | 12.87% | 39 | 6.34% |
| Public Administration and Defence | 29 | 2.57% | 1 | 0.19% | 28 | 4.55% |
| Transportation and Storage | 38 | 3.37% | 10 | 1.95% | 28 | 4.55% |
| Wholesale and Retail Trade | 63 | 5.59% | 15 | 2.92% | 48 | 7.8% |

## Table 2: Immigrant Social Integration and Commuting Distance of Home Renters

| | (1) | (2) | (3) | (4) | (5) | (6) | (7) | (8) |
|---|---|---|---|---|---|---|---|---|
| | Y: log (Commuting Distance) | | | | | | | |
| | OLS | OLS | Main IV | Alternative IV | OLS | OLS | OLS | OLS |
| Immigration Years | 0.0146*** | 0.0120** | 0.0304** | 0.0280** | | | | |
| | (0.0048) | (0.0052) | (0.0152) | (0.0143) | | | | |
| Attend Community Events | | | | | 0.0760** | 0.0881** | | |
| | | | | | (0.0344) | (0.0390) | | |
| Subjective Integration | | | | | | | 0.0570 | 0.0029 |
| | | | | | | | (0.0394) | (0.0404) |
| Age | | 0.0009 | -0.0050 | -0.0032 | | 0.0068 | | 0.0059 |
| | | (0.0070) | (0.0072) | (0.0069) | | (0.0065) | | (0.0066) |
| Gender | | -0.0599 | -0.0440 | -0.0448 | | -0.0533 | | -0.0632 |
| | | (0.0968) | (0.0870) | (0.0863) | | (0.0958) | | (0.0953) |
| Married | | 0.0185 | 0.0185 | 0.0247 | | -0.0143 | | 0.0030 |
| | | (0.1041) | (0.1156) | (0.1137) | | (0.1049) | | (0.1038) |
| With Child | | 0.0078 | 0.0356 | 0.0369 | | -0.0094 | | -0.0033 |
| | | (0.1002) | (0.0937) | (0.0925) | | (0.0992) | | (0.1006) |
| Income (S$2,000-S$4,000) | | 0.4737* | 0.5281 | 0.5164 | | 0.4420 | | 0.4330 |
| | | (0.2844) | (0.3727) | (0.3738) | | (0.2835) | | (0.2854) |
| Income (S$4,000-S$8,000) | | 0.5699** | 0.5902* | 0.5903* | | 0.5268** | | 0.5469** |
| | | (0.2207) | (0.3154) | (0.3142) | | (0.2199) | | (0.2207) |
| Income (over S$8,000) | | 0.6213** | 0.6935*** | 0.6665** | | 0.5359** | | 0.5759** |
| | | (0.2447) | (0.2631) | (0.2641) | | (0.2447) | | (0.2435) |
| Ever Moved Home | | 0.1361 | 0.0745 | 0.0843 | | 0.1927* | | 0.1756 |
| | | (0.1088) | (0.1139) | (0.1104) | | (0.1092) | | (0.1100) |
| Race (Indian) | | -0.0249 | -0.0268 | -0.0244 | | -0.0815 | | -0.0349 |
| | | (0.1013) | (0.0978) | (0.0956) | | (0.1060) | | (0.1120) |
| Race (Malay) | | -0.0095 | 0.1239 | 0.1199 | | 0.0321 | | 0.0179 |
| | | (0.2405) | (0.2684) | (0.2792) | | (0.2503) | | (0.2515) |
| Race (Others) | | 0.0220 | 0.0153 | 0.0283 | | 0.0132 | | 0.0222 |
| | | (0.0972) | (0.1198) | (0.1162) | | (0.0964) | | (0.0969) |
| Occupation Controls | N | Y | Y | Y | N | Y | N | Y |
| Time Fixed Effect | Y | Y | Y | Y | Y | Y | Y | Y |
| District Fixed Effect | Y | Y | Y | Y | Y | Y | Y | Y |
| First-Stage F-stats | | | 152.17 | 155.82 | | | | |
| Observations | 512 | 512 | 489 | 498 | 512 | 512 | 512 | 512 |
| R-squared | 0.268 | 0.345 | 0.341 | 0.340 | 0.264 | 0.346 | 0.262 | 0.341 |

Note: Columns (1)-(2) report the OLS estimation result using immigration years as the independent variable. Column (3) reports the result using the growth rate of the foreign workforce as the main IV. Column (4) reports the result using the growth rate of the foreign population as the alternative IV. In Columns (3) and (4), the numbers of observations of the IV estimations are smaller due to missing information on the foreign workforce and population in the early years. Columns (5)-(6) report OLS estimation results using frequency of attendance at community activities as the independent variables, and Columns (7)-(8) report OLS estimation results using the subjective score of integration as the independent variables. Occupation control includes a series of dummies representing occupations. All variables are defined in Online Appendix Table 1. Standard errors are clustered at building level for OLS estimations and at arrival-year level for IV estimations. Robust standard errors in parentheses. *** $p<0.01$, ** $p<0.05$, * $p<0.1$

## Table 3: Immigrant Social Integration and Housing Rent

| | (1) | (2) | (3) | (4) | (5) | (6) | (7) | (8) |
|---|---|---|---|---|---|---|---|---|
| | Y: log (Rent) | | | | | | | |
| | OLS | OLS | Main IV | Alternative IV | OLS | OLS | OLS | OLS |
| Immigration Years | -0.0013* | -0.0022** | -0.0067*** | -0.0056*** | | | | |
| | (0.0004) | (0.0010) | (0.0025) | (0.0022) | | | | |
| Attend Community Events | | | | | -0.0139* | -0.0108* | | |
| | | | | | (0.0078) | (0.0063) | | |
| Subjective Integration | | | | | | | -0.0180** | -0.0123** |
| | | | | | | | (0.0076) | (0.0062) |
| Floor | | 0.0040*** | 0.0039*** | 0.0038** | | 0.0041*** | | 0.0041*** |
| | | (0.0012) | (0.0015) | (0.0015) | | (0.0012) | | (0.0012) |
| Building Age | | -0.0034*** | -0.0032*** | -0.0033*** | | -0.0035*** | | -0.0035*** |
| | | (0.0008) | (0.0008) | (0.0007) | | (0.0008) | | (0.0008) |
| log (Distance to CBD) | | -0.1131 | -0.0640 | -0.0966 | | -0.1314* | | -0.1271* |
| | | (0.0730) | (0.0744) | (0.0727) | | (0.0725) | | (0.0714) |
| log (Distance to MRT) | | -0.0068 | -0.0264*** | -0.0087 | | -0.0065 | | -0.0042 |
| | | (0.0198) | (0.0080) | (0.0174) | | (0.0200) | | (0.0197) |
| log (Distance to Bus Stop) | | -0.0003 | -0.0011 | 0.0027 | | -0.0045 | | -0.0033 |
| | | (0.0110) | (0.0086) | (0.0082) | | (0.0112) | | (0.0112) |
| log (Distance to Hospital) | | -0.0353* | -0.0268** | -0.0324*** | | -0.0333* | | -0.0342* |
| | | (0.0184) | (0.0122) | (0.0113) | | (0.0183) | | (0.0179) |
| log (Distance to Park) | | 0.0014 | 0.0026 | 0.0012 | | 0.0023 | | 0.0021 |
| | | (0.0082) | (0.0063) | (0.0061) | | (0.0082) | | (0.0080) |
| Executive Condominium | | 0.2040** | 0.2279*** | 0.2004*** | | 0.2099** | | 0.1970** |
| | | (0.0848) | (0.0551) | (0.0584) | | (0.0870) | | (0.0872) |
| Demographics Control | N | Y | Y | Y | N | Y | N | Y |
| Time Fixed Effect | Y | Y | Y | Y | Y | Y | Y | Y |
| District Fixed Effect | Y | Y | Y | Y | Y | Y | Y | Y |
| First-Stage F-stats | | | 53.76 | 39.71 | | | | |
| Observations | 435 | 435 | 401 | 431 | 435 | 435 | 435 | 435 |
| R-squared | 0.245 | 0.618 | 0.640 | 0.610 | 0.248 | 0.616 | 0.253 | 0.617 |

Note: Samples that did not rent a whole flat are excluded. Columns (1)-(2) report the OLS estimation result using immigration years as the independent variable. Column (3) reports the result using the growth rate of the foreign workforce as the main IV. Column (4) reports the result using the growth rate of the foreign population as the alternative IV. In Columns (3) and (4), the numbers of observations of the IV estimations are smaller due to missing information on the foreign workforce and population in the early years. Columns (5)-(6) report OLS estimation results using frequency of attendance at community activities as the independent variables, and Columns (7)-(8) report OLS estimation results using the subjective score of integration as the independent variables. Unreported demographic controls are same as in Table 2, which include age, gender, marital status, income, race, with child dummy, and ever moved home dummy. All variables are defined in Online Appendix Table 1. Standard errors are clustered at building level for OLS estimations and at arrival-year level for IV estimations. Robust standard errors in parentheses. *** p<0.01, ** p<0.05, * p<0.1

**Table 4: Immigrant Social Integration and Commuting Distance of Homeowners**

| | (1) | (2) | (3) | (4) |
|---|---|---|---|---|
| | Y: log (Price) | | | |
| | Naturalised & Locally Born Citizens | Naturalised & Locally Born Citizens | Naturalised & Locally Born Citizens | Naturalised & Locally Born Citizens |
| | OLS | OLS | OLS | OLS |
| Attend Community Events × Naturalised Citizen | -0.0497 | -0.1591 | | |
| | (0.1800) | (0.1594) | | |
| Attend Community Events | 0.0338 | 0.0793 | | |
| | (0.1260) | (0.1123) | | |
| Subjective Integration × Naturalised Citizen | | | -0.2001 | 0.0128 |
| | | | (0.2083) | (0.1947) |
| Subjective Integration | | | 0.0696 | 0.0268 |
| | | | (0.0998) | (0.1190) |
| Naturalised Citizen | -0.0131 | -0.0722 | -0.0129 | -0.1036 |
| | (0.2074) | (0.2464) | (0.1957) | (0.2463) |
| Demographics Control | N | Y | N | Y |
| Occupation Fixed Effect | N | Y | N | Y |
| District Fixed Effect | Y | Y | Y | Y |
| Time Fixed Effect | Y | Y | Y | Y |
| Observations | 413 | 413 | 413 | 413 |
| R-squared | 0.068 | 0.156 | 0.070 | 0.154 |

Note: Regression sample includes locally born and naturalised citizen homeowners in the Singapore public resale housing market. Columns (1)-(2) report the OLS estimation results using the frequency of attending community events as the explanatory variable, and Columns (3)-(4) report the OLS estimation results using the subjective score of integration as the explanatory variable. Occupation control includes a series of dummies representing occupations. Unreported demographic controls are same as in Table 2, which include age, gender, marital status, income, race, with child dummy, and ever moved home dummy. All variables are defined in Online Appendix Table 1. Standard errors clustered at building level. Robust standard errors in parentheses. *** p<0.01, ** p<0.05, * p<0.1

**Table 5: Immigrant Social Integration and Housing Price**

| | (1) | (2) | (3) | (4) |
|---|---|---|---|---|
| | Y: log (Price) | | | |
| | Naturalised & Locally Born Citizens | Naturalised & Locally Born Citizens | Naturalised & Locally Born Citizens | Naturalised & Locally Born Citizens |
| | OLS | OLS | OLS | OLS |
| Attend Community Events × Naturalised Citizen | 0.0067 | 0.0094 | | |
| | (0.0138) | (0.0103) | | |
| Attend Community Events | -0.0086 | -0.0081 | | |
| | (0.0064) | (0.0061) | | |
| Subjective Integration × Naturalised Citizen | | | -0.0099 | -0.0144 |
| | | | (0.0151) | (0.0107) |
| Subjective Integration | | | 0.0093 | 0.0083 |
| | | | (0.0073) | (0.0058) |
| Naturalised Citizen | -0.0015 | -0.0137 | -0.0021 | -0.0131 |
| | (0.0178) | (0.0136) | (0.0184) | (0.0139) |
| Housing Features Control | N | Y | N | Y |
| Demographics Control | N | Y | N | Y |
| District Fixed Effect | Y | Y | Y | Y |
| Time Fixed Effect | Y | Y | Y | Y |
| Observations | 413 | 413 | 413 | 413 |
| R-squared | 0.747 | 0.875 | 0.747 | 0.875 |

Note: Regression sample includes locally born and naturalised citizen homeowners in the Singapore public resale housing market. Columns (1)-(2) report the OLS estimation results using the frequency of attending community events as the explanatory variable, and Columns (3)-(4) report the OLS estimation results using the subjective score of integration as the explanatory variable. Unreported housing features control are same as in Table 3, which include floor, building age, distances to CBD, MTR stations, bus stops, hospitals and green parts, as well as the dummy variable of executive condominiums. Unreported demographic controls are same as in Table 2, which include age, gender, marital status, income, race, with child dummy, and ever moved home dummy. All variables are defined in Online Appendix Table 1. Standard errors clustered at building level. Robust standard errors in parentheses. *** $p<0.01$, ** $p<0.05$, * $p<0.1$

# Social Integration and Housing Behaviours of Immigrants:

# Evidence from Singapore's Public Housing Market

## Online Appendix

## Appendix A. Supplementary Tables

**Appendix Table 1: Definition of Variables**

| Variable Name | Definition |
|---|---|
| log (Commuting Distance) | Commuting distance of a renter from home to workplace, in logarithmic form (km). |
| Subjective Integration | Standardised score for the question that asks respondents to rank their sense of belonging to Singapore on a 5-point scale. |
| Attend Community Events | Aggregate and standardised score for survey questions related to the individual's objective attendance at various community events: voluntary work, festival celebrations, and policy dialogues. |
| log (Rent) | Monthly rent of the unit in logarithmic form (S$). |
| log (Price) | Purchase price of the unit in logarithmic form (S$). |
| Migration Years | Years of post-migration residency in Singapore for the respondent who is not locally born. |
| Earlier Immigrants | A dummy variable equal to 1 if an immigrant has resided for over 10 years, zero otherwise. |
| Immigration Years at Purchase | Years of post-immigration residency in Singapore for the respondent when purchasing the property, excluding the locally born citizens. |
| Immigrant | 1 = Yes<br>0 = No |
| Naturalised Citizen | 1 = Yes<br>0 = No |
| Age | Age of the respondent (years). |
| Owner | Whether the respondent is an owner of the public housing unit:<br>1 = Yes<br>0 = No |
| Gender | 1 = Male<br>0 = Female |
| Race | A series of dummy variables representing the race of the respondent:<br>1 = Chinese<br>2 = Indian<br>3 = Malay<br>4 = Other |
| Married | 1 = Yes (married)<br>0 = No (never married/divorced/separated) |
| With Child | 1 = Yes (have at least one child)<br>0 = No |
| Nationality | Series of dummy variables representing the nationality of the respondent:<br>1 = Foreigner<br>2 = Singapore PR<br>3 = Naturalised to Singapore citizenship after birth<br>4 = Born with Singapore citizenship |
| Education Level | Series of dummy variables representing the education level of the respondent:<br>1 = Secondary school or below<br>2 = Junior college/polytechnic<br>3 = Bachelor's degree<br>4 = Postgraduate degree |
| Income | Series of dummy variables (Income (S$2,000S$4,000); Income (S$4,000-S$8,000); Income (over S$8,000)) representing the income level of the respondent:<br>1 = Below S$2,000<br>2 = S$2,001-S$4,000<br>3 = S$4,001-S$8,000<br>4 = Above S$8,000 |
| Ever Moved Home | Whether the respondent has ever moved home after immigration to Singapore:<br>1 = Yes<br>0 = No |
| HDB Town | A series of dummy variables representing the district:<br>1 = Bukit Panjang<br>2 = Clementi<br>3 = Queenstown |
| Property Type | A series of dummy variables representing the property type:<br>1 = HDB 1-Room Unit<br>2 = HDB 2-Room Unit<br>3 = HDB 3-Room Unit<br>4 = HDB 4-Room Unit<br>5 = HDB 5-Room Unit<br>6 = HDB Executive Condominium |
| Executive Condominium | Whether the property is an executive condominium:<br>1 = Yes<br>0 = No |
| Occupation Industry | Series of dummy variables representing the occupation of the respondent:<br>1 = Accommodation and Food Services Activities<br>2 = Arts, Entertainment and Recreation<br>3 = Education<br>4 = Electricity, Gas, Steam and Air-Conditioning Supply, Water Supply<br>5 = Financial, Real Estate and Insurance Activities<br>6 = Health and Social Services<br>7 = Information and Communications<br>8 = Manufacturing and Construction<br>9 = Other<br>10 = Professional, Scientific and Technical Activities<br>11 = Public Administration and Defense<br>12 = Transportation and Storage<br>13 = Wholesale and Retail Trade |
| Rent a Room in a Flat | Whether the respondent only sublets one room (or some rooms) in the unit, instead of renting the entire unit:<br>1 = Yes<br>0 = No |
| Contract Year | For renters, the is the year the current rental contract was signed. For owners, this is the year the unit was purchased. |
| Floor | Floor of the unit. |
| Building Age | Building age of the unit. |
| Unit Size | Gross floor area of the unit. |
| log (Distance to CBD) | Distance from the unit to the CBD of Singapore, in logarithmic form (km). |
| log (Distance to MRT) | Distance from the unit to the closet MRT train station, in logarithmic form (km). |
| log (Distance to Bus Stop) | Distance from the unit to the closet bus stop, in logarithmic form (km). |
| log (Distance to Hospital) | Distance from the unit to the closet hospital, in logarithmic form k(m). |
| log (Distance to Park) | Distance from the unit to the closet park, in logarithmic form (km). |

**Appendix Table 2: VIF Test Results for OLS Models in Table 2 - Immigrant Social Integration and Commuting Distance of Home Renters**

| | (1)<br>OLS | (2)<br>OLS | (5)<br>OLS | (6)<br>OLS | (7)<br>OLS | (8)<br>OLS |
|---|---|---|---|---|---|---|
| Immigration Years | 1.017 | 1.589 | | | | |
| Attend Community Events | | | 1.011 | 1.183 | | |
| Subjective Integration | | | | | 1.039 | 1.368 |
| Age | | 1.983 | | 1.646 | | 1.662 |
| Gender | | 1.199 | | 1.202 | | 1.203 |
| Married | | 1.586 | | 1.585 | | 1.577 |
| With Child | | 1.097 | | 1.092 | | 1.093 |
| Income (S$2,000-S$4,000) | | 3.012 | | 2.982 | | 2.981 |
| Income (S$4,000-S$8,000) | | 4.357 | | 4.349 | | 4.376 |
| Income (over S$8,000) | | 3.330 | | 3.309 | | 3.292 |
| Ever Moved Home | | 1.249 | | 1.175 | | 1.181 |
| Race (Indian) | | 1.757 | | 1.848 | | 1.912 |
| Race (Malay) | | 1.144 | | 1.142 | | 1.143 |
| Race (Others) | | 1.658 | | 1.661 | | 1.663 |

**Appendix Table 3: VIF Test Results for OLS Models in Table 3 - Immigrant Social Integration and Housing Rent**

| | (1) OLS | (2) OLS | (5) OLS | (6) OLS | (7) OLS | (8) OLS |
|---|---|---|---|---|---|---|
| Immigration Years | 1.021 | 1.701 | | | | |
| Attend Community Events | | | 1.023 | 1.269 | | |
| Subjective Integration | | | | | 1.050 | 1.417 |
| Floor | | 1.343 | | 1.341 | | 1.341 |
| Building Age | | 3.631 | | 3.632 | | 3.647 |
| log (Distance to CBD) | | 1.109 | | 1.349 | | 1.175 |
| log (Distance to MRT) | | 2.347 | | 2.349 | | 2.340 |
| log (Distance to Bus Stop) | | 1.263 | | 1.259 | | 1.249 |
| log (Distance to Hospital) | | 1.586 | | 1.554 | | 1.562 |
| log (Distance to Park) | | 1.340 | | 1.346 | | 1.342 |
| Executive Condominium | | 5.464 | | 5.466 | | 5.494 |
| Age | | 2.228 | | 1.831 | | 1.846 |
| Gender | | 1.250 | | 1.250 | | 1.253 |
| Married | | 2.170 | | 2.169 | | 2.165 |
| With Child | | 1.799 | | 1.797 | | 1.797 |
| Income (S$2,000-S$4,000) | | 3.204 | | 3.119 | | 3.121 |
| Income (S$4,000-S$8,000) | | 4.941 | | 4.883 | | 4.914 |
| Income (over S$8,000) | | 3.914 | | 3.829 | | 3.824 |
| Ever Moved Home | | 1.404 | | 1.334 | | 1.331 |
| Race (Indian) | | 1.957 | | 2.049 | | 2.107 |
| Race (Malay) | | 1.184 | | 1.179 | | 1.179 |
| Race (Others) | | 1.755 | | 1.760 | | 1.755 |

**Appendix Table 4: VIF Test Results for OLS Models in Table 4 - Immigrant Social Integration and Commuting Distance of Homeowners**

| | (1) OLS | (2) OLS | (5) OLS | (6) OLS | (7) OLS | (8) OLS |
|---|---|---|---|---|---|---|
| Attend Community Events × Naturalised Citizen | 1.420 | 1.516 | | | 1.420 | 1.516 |
| Attend Community Events | 1.403 | 1.555 | | | 1.403 | 1.555 |
| Subjective Integration × Naturalised Citizen | | | 1.401 | 1.491 | | |
| Subjective Integration | | | 1.411 | 1.539 | | |
| Naturalised Citizen | 1.111 | 1.512 | 1.106 | 1.500 | 1.111 | 1.512 |
| Age | | 1.440 | | 1.452 | | 1.440 |
| Gender | | 1.317 | | 1.316 | | 1.317 |
| Married | | 1.264 | | 1.267 | | 1.264 |
| With Child | | 1.217 | | 1.205 | | 1.217 |
| Income (S$2,000-S$4,000) | | 3.217 | | 3.035 | | 3.217 |
| Income (S$4,000-S$8,000) | | 1.484 | | 1.331 | | 1.484 |
| Income (over S$8,000) | | 4.938 | | 4.881 | | 4.938 |
| Ever Moved Home | | 1.196 | | 1.186 | | 1.196 |
| Race (Indian) | | 1.235 | | 1.254 | | 1.235 |
| Race (Malay) | | 1.278 | | 1.272 | | 1.278 |
| Race (Others) | | 1.259 | | 1.257 | | 1.259 |

**Appendix Table 5: VIF Test Results for OLS Models in Table 5 - Immigrant Social Integration and Housing Price**

| | (1)<br>OLS | (2)<br>OLS | (5)<br>OLS | (6)<br>OLS | (7)<br>OLS | (8)<br>OLS |
|---|---|---|---|---|---|---|
| Attend Community Events × Naturalised Citizen | 1.413 | 1.522 | | | 1.413 | 1.522 |
| Attend Community Events | 1.458 | 1.570 | | | 1.458 | 1.570 |
| Subjective Integration × Naturalised Citizen | | | 1.348 | 1.466 | | |
| Subjective Integration | | | 1.370 | 1.480 | | |
| Naturalised Citizen | 1.124 | 1.528 | 1.105 | 1.520 | 1.124 | 1.528 |
| Floor | | 2.039 | | 2.035 | | 2.039 |
| Building Age | | 1.593 | | 1.577 | | 1.593 |
| log (Distance to CBD) | | 4.484 | | 4.479 | | 4.484 |
| log (Distance to MRT) | | 3.498 | | 3.523 | | 3.498 |
| log (Distance to Bus Stop) | | 2.117 | | 2.117 | | 2.117 |
| log (Distance to Hospital) | | 1.273 | | 1.275 | | 1.273 |
| log (Distance to Park) | | 3.733 | | 3.958 | | 3.733 |
| Executive Condominium | | 1.549 | | 1.559 | | 1.549 |
| Age | | 1.681 | | 1.706 | | 1.681 |
| Gender | | 1.209 | | 1.211 | | 1.209 |
| Married | | 1.292 | | 1.276 | | 1.292 |
| With Child | | 1.219 | | 1.212 | | 1.219 |
| Income (S$2,000-S$4,000) | | 1.837 | | 1.828 | | 1.837 |
| Income (S$4,000-S$8,000) | | 2.352 | | 2.397 | | 2.352 |
| Income (over S$8,000) | | 1.920 | | 1.907 | | 1.920 |
| Ever Moved Home | | 1.213 | | 1.238 | | 1.213 |
| Race (Indian) | | 1.299 | | 1.300 | | 1.299 |
| Race (Malay) | | 1.430 | | 1.462 | | 1.430 |
| Race (Others) | | 1.282 | | 1.288 | | 1.282 |

**Appendix Table 6: Immigrant Social Integration and Commuting Distance—First-stage Result of IV Estimations**

| | (1) | (2) |
|---|---|---|
| | Y: Immigration Years | |
| | Main IV | Alternative IV |
| Growth Rate of Foreign Workforce | 42.3357*** | |
| | (4.1147) | |
| Growth Rate of Foreign Population | | 56.4093*** |
| | | (5.7109) |
| Demographics Control | Y | Y |
| Labor Market Control | Y | Y |
| Occupation Control | Y | Y |
| District Fixed Effect | Y | Y |
| First-Stage F-stats | 152.17 | 155.82 |
| Observations | 489 | 498 |

Note: Column (1) reports the first-stage result using growth rate of foreign workforce as the main IV. Column (2) reports the first-stage result using growth rate of foreign population as the alternative IV. Dropping data in IV estimations due to missing information on foreign workforce and population in the early years. Definitions of the other variables are reported in Appendix Table 1. Standard errors clustered at the arrival year level. Robust standard errors in parentheses. *** p<0.01, ** p<0.05, * p<0.1

## Appendix Table 7: Robustness Check for Immigrant Social Integration and Commuting Distance of Home Renters Using District × Time Fixed Effects

| | (1) | (2) | (3) | (4) | (5) | (6) | (7) | (8) |
|---|---|---|---|---|---|---|---|---|
| | Y: log (Commuting Distance) | | | | | | | |
| | OLS | OLS | Main IV | Alternative IV | OLS | OLS | OLS | OLS |
| Immigration Years | 0.0148*** | 0.0123** | 0.0300** | 0.0276** | | | | |
| | (0.0048) | (0.0053) | (0.0147) | (0.0139) | | | | |
| Attend Community Events | | | | | 0.0760** | 0.0861** | | |
| | | | | | (0.0343) | (0.0386) | | |
| Subjective Integration | | | | | | | 0.0561 | 0.0036 |
| | | | | | | | (0.0396) | (0.0406) |
| Demographic Controls | N | Y | Y | Y | N | Y | N | Y |
| Occupation Controls | N | Y | Y | Y | N | Y | N | Y |
| District×Time Fixed Effect | Y | Y | Y | Y | Y | Y | Y | Y |
| First-Stage F-stats | | | 153.07 | 156.77 | | | | |
| Observations | 512 | 512 | 489 | 498 | 512 | 512 | 512 | 512 |
| R-squared | 0.268 | 0.345 | 0.341 | 0.340 | 0.264 | 0.346 | 0.262 | 0.341 |

Note: Columns (1)-(2) report the OLS estimation result using immigration years as the independent variable. Column (3) reports the result using the growth rate of the foreign workforce as the main IV. Column (4) reports the result using the growth rate of the foreign population as the alternative IV. In Columns (3) and (4), the numbers of observations of the IV estimations are smaller due to missing information on the foreign workforce and population in the early years. Columns (5)-(6) report OLS estimation results using frequency of attendance at community activities as the independent variables, and Columns (7)-(8) report OLS estimation results using the subjective score of integration as the independent variables. Occupation control includes a series of dummies representing occupations. All variables are defined in Online Appendix Table 1. Standard errors are clustered at building level for OLS estimations and at arrival-year level for IV estimations. Robust standard errors in parentheses. *** p<0.01, ** p<0.05, * p<0.1

## Appendix Table 8: Robustness Check for Immigrant Social Integration and Housing Rent Using District × Time Fixed Effects

| | (1) | (2) | (3) | (4) | (5) | (6) | (7) | (8) |
|---|---|---|---|---|---|---|---|---|
| | Y: log (Rent) | | | | | | | |
| | OLS | OLS | Main IV | Alternative IV | OLS | OLS | OLS | OLS |
| Immigration Years | -0.0015* | -0.0025*** | -0.0064*** | -0.0058*** | | | | |
| | (0.0005) | (0.0010) | (0.0024) | (0.0021) | | | | |
| Attend Community Events | | | | | -0.0139* | -0.0103* | | |
| | | | | | (0.0078) | (0.0062) | | |
| Subjective Integration | | | | | | | -0.0180** | -0.0123** |
| | | | | | | | (0.0076) | (0.0062) |
| Demographics Control | N | Y | Y | Y | N | Y | N | Y |
| Building Feature Control | N | Y | Y | Y | N | Y | N | Y |
| District×Time Fixed Effect | Y | Y | Y | Y | Y | Y | Y | Y |
| First-Stage F-stats | | | 56.74 | 42.02 | | | | |
| Observations | 435 | 435 | 401 | 431 | 435 | 435 | 435 | 435 |
| R-squared | 0.259 | 0.624 | 0.648 | 0.616 | 0.248 | 0.621 | 0.253 | 0.622 |

Note: Samples that did not rent a whole flat are excluded. Columns (1)-(2) report the OLS estimation result using immigration years as the independent variable. Column (3) reports the result using the growth rate of the foreign workforce as the main IV. Column (4) reports the result using the growth rate of the foreign population as the alternative IV. In Columns (3) and (4), the numbers of observations of the IV estimations are smaller due to missing information on the foreign workforce and population in the early years. Columns (5)-(6) report OLS estimation results using frequency of attendance at community activities as the independent variables, and Columns (7)-(8) report OLS estimation results using the subjective score of integration as the independent variables. Unreported demographic controls are same as in Table 2, which include age, gender, marital status, income, race, with child dummy, and ever moved home dummy. All variables are defined in Online Appendix Table 1. Standard errors are clustered at building level for OLS estimations and at arrival-year level for IV estimations. Robust standard errors in parentheses. *** p<0.01, ** p<0.05, * p<0.1

## Appendix Table 9: Robustness Check for Immigrant Social Integration and Commuting Distance of Home Renters Using Wald Bootstrap Inference

| | (1) | (2) | (3) | (4) | (5) | (6) | (7) | (8) |
|---|---|---|---|---|---|---|---|---|
| | Y: log (Commuting Distance) | | | | | | | |
| | OLS | OLS | Main IV | Alternative IV | OLS | OLS | OLS | OLS |
| Immigration Years | 0.0146*** | 0.0120** | 0.0304* | 0.0280* | | | | |
| Attend Community Events | | | | | 0.0760** | 0.0881** | | |
| Subjective Integration | | | | | | | 0.0570 | 0.0029 |
| z-score | 3.3340 | 2.2329 | 1.8198 | 1.6890 | 2.2111 | 2.2609 | 1.4455 | 0.0712 |
| p-value | 0.0000 | 0.0320 | 0.0720 | 0.0880 | 0.0260 | 0.0300 | 0.1780 | 0.9520 |

Note: This table report the robustness check results for the estimation results in Table 2, using Wald cluster bootstrap method to calculate the statistical power. The replication number equals 500. *** p<0.01, ** p<0.05, * p<0.1

## Appendix Table 10: Robustness Check for Immigrant Social Integration and Housing Rent Using Wald Bootstrap Inference

| | (1) | (2) | (3) | (4) | (5) | (6) | (7) | (8) |
|---|---|---|---|---|---|---|---|---|
| | Y: log (Commuting Distance) | | | | | | | |
| | OLS | OLS | Main IV | Alternative IV | OLS | OLS | OLS | OLS |
| Immigration Years | -0.0013 | -0.0022** | -0.0067** | -0.0056** | | | | |
| Attend Community Events | | | | | -0.0139* | -0.0108* | | |
| Subjective Integration | | | | | | | -0.0180** | -0.0123** |
| z-score | -1.5397 | -2.1829 | -2.3780 | -2.4139 | -1.7809 | -1.7278 | -2.3698 | -1.9868 |
| p-value | 0.1360 | 0.0160 | 0.0260 | 0.0140 | 0.0780 | 0.0740 | 0.0200 | 0.0420 |

Note: This table report the robustness check results for the estimation results in Table 3, using Wald cluster bootstrap method to calculate the statistical power. The replication number equals 500. *** p<0.01, ** p<0.05, * p<0.1

**Appendix Table 11: Difference between the Commuting Distances of Home Renters and Homeowners**

| | (1) |
|---|---|
| | Y: log (Commuting Distance) |
| | OLS |
| Renter | 0.2227*** |
| | (0.0800) |
| Age | -0.0005 |
| | (0.0042) |
| Gender | -0.0755 |
| | (0.0684) |
| Married | 0.0070 |
| | (0.0802) |
| With Child | -0.0742 |
| | (0.0696) |
| Income (S$2,000-S$4,000) | 0.6436** |
| | (0.2781) |
| Income (S$4,000-S$8,000) | 0.8038*** |
| | (0.2558) |
| Income (over S$8,000) | 0.6151** |
| | (0.2882) |
| Ever Moved Home | 0.1060 |
| | (0.0754) |
| Race (Indian) | -0.0793 |
| | (0.0800) |
| Race (Malay) | 0.0230 |
| | (0.1445) |
| Race (Others) | -0.1612 |
| | (0.1009) |
| Occupation Controls | Y |
| Time Fixed Effect | Y |
| District Fixed Effect | Y |
| First-Stage F-stats | |
| Observations | 925 |
| R-squared | 0.222 |

Note: Regression sample includes home renters in the Singapore public rental housing market, and the locally born and naturalised citizen homeowners in the Singapore public resale housing market. *Renter* is a dummy variable indicating the home renters. Occupation control includes a series of dummies representing occupations. All variables are defined in Online Appendix Table 1. Standard errors clustered at building level. Robust standard errors in parentheses. *** p<0.01, ** p<0.05, * p<0.1

## Appendix Table 12: Heterogeneity Analysis for Immigrant Social Integration and Commuting Distance of Home Renters

### Panel A. Non-Chinese Renters

| | (1) | (2) | (3) | (4) | (5) | (6) | (7) | (8) |
|---|---|---|---|---|---|---|---|---|
| | Y: log (Commuting Distance) | | | | | | | |
| | OLS | OLS | Main IV | Alternative IV | OLS | OLS | OLS | OLS |
| Immigration Years | 0.0164*** | 0.0182** | 0.0601*** | 0.0553*** | | | | |
| | (0.0060) | (0.0073) | (0.0148) | (0.0153) | | | | |
| Attend Community Events | | | | | 0.1109*** | 0.1294*** | | |
| | | | | | (0.0400) | (0.0439) | | |
| Subjective Integration | | | | | | | 0.0476 | 0.0089 |
| | | | | | | | (0.0484) | (0.0436) |
| Demographic Controls | N | Y | Y | Y | N | Y | N | Y |
| Occupation Controls | N | Y | Y | Y | N | Y | N | Y |
| Time Fixed Effect | Y | Y | Y | Y | Y | Y | Y | Y |
| District Fixed Effect | Y | Y | Y | Y | Y | Y | Y | Y |
| First-Stage F-stats | | | 74.67 | 105.70 | | | | |
| Observations | 355 | 355 | 340 | 347 | 355 | 355 | 355 | 355 |
| R-squared | 0.286 | 0.371 | 0.368 | 0.367 | 0.285 | 0.376 | 0.276 | 0.363 |

### Panel B. Chinese Renters

| | (1) | (2) | (3) | (4) | (5) | (6) | (7) | (8) |
|---|---|---|---|---|---|---|---|---|
| | Y: log (Commuting Distance) | | | | | | | |
| | OLS | OLS | Main IV | Alternative IV | OLS | OLS | OLS | OLS |
| Immigration Years | 0.0104 | 0.0053 | -0.0104 | -0.0185 | | | | |
| | (0.0099) | (0.0120) | (0.0226) | (0.0226) | | | | |
| Attend Community Events | | | | | -0.0702 | -0.0433 | | |
| | | | | | (0.0937) | (0.1077) | | |
| Subjective Integration | | | | | | | 0.0802 | 0.0174 |
| | | | | | | | (0.0685) | (0.0867) |
| Demographic Controls | N | Y | Y | Y | N | Y | N | Y |
| Occupation Controls | N | Y | Y | Y | N | Y | N | Y |
| Time Fixed Effect | Y | Y | Y | Y | Y | Y | Y | Y |
| District Fixed Effect | Y | Y | Y | Y | Y | Y | Y | Y |
| First-Stage F-stats | | | 108.07 | 69.39 | | | | |
| Observations | 157 | 157 | 149 | 151 | 157 | 157 | 157 | 157 |
| R-squared | 0.240 | 0.353 | 0.379 | 0.369 | 0.239 | 0.353 | 0.241 | 0.352 |

Note: Panels A and B replicate the analyses in Table 2 using subsamples of Chinese and non-Chinese migrant renters, respective. Columns (1)-(2) report the OLS estimation result using immigration years as the independent variable. Column (3) reports the result using the growth rate of the foreign workforce as the main IV. Column (4) reports the result using the growth rate of the foreign population as the alternative IV. In Columns (3) and (4), the numbers of observations of the IV estimations are smaller due to missing information on the foreign workforce and population in the early years. Columns (5)-(6) report OLS estimation results using frequency of attendance at community activities as the independent variables, and Columns (7)-(8) report OLS estimation results using the subjective score of integration as the independent variables. Occupation control includes a series of dummies representing occupations. All variables are defined in Online Appendix Table 1. Standard errors are clustered at building level for OLS estimations and at arrival-year level for IV estimations. Robust standard errors in parentheses. *** p<0.01, ** p<0.05, * p<0.1

## Appendix Table 13: Heterogeneity Analysis for Immigrant Social Integration and Housing Rent

### Panel A. Non-Chinese Renters

| | (1) | (2) | (3) | (4) | (5) | (6) | (7) | (8) |
|---|---|---|---|---|---|---|---|---|
| | Y: log (Rent) | | | | | | | |
| | OLS | OLS | Main IV | Alternative IV | OLS | OLS | OLS | OLS |
| Immigration Years | -0.0018* | -0.0031** | -0.0055** | -0.0050** | | | | |
| | (0.0005) | (0.0013) | (0.0024) | (0.0025) | | | | |
| Attend Community Events | | | | | -0.0099 | -0.0131* | | |
| | | | | | (0.0081) | (0.0067) | | |
| Subjective Integration | | | | | | | -0.0147* | -0.0153** |
| | | | | | | | (0.0087) | (0.0062) |
| Demographics Control | N | Y | Y | Y | N | Y | N | Y |
| Building Feature Control | N | Y | Y | Y | N | Y | N | Y |
| Time Fixed Effect | Y | Y | Y | Y | Y | Y | Y | Y |
| District Fixed Effect | Y | Y | Y | Y | Y | Y | Y | Y |
| First-Stage F-stats | | | 55.52 | 59.69 | | | | |
| Observations | 303 | 303 | 286 | 300 | 303 | 303 | 303 | 303 |
| R-squared | 0.241 | 0.619 | 0.665 | 0.621 | 0.239 | 0.616 | 0.244 | 0.618 |

### Panel B. Chinese Renters

| | (1) | (2) | (3) | (4) | (5) | (6) | (7) | (8) |
|---|---|---|---|---|---|---|---|---|
| | Y: log (Rent) | | | | | | | |
| | OLS | OLS | Main IV | Alternative IV | OLS | OLS | OLS | OLS |
| Immigration Years | -0.0010 | -0.0026 | -0.0111 | -0.0065 | | | | |
| | (0.0023) | (0.0026) | (0.0108) | (0.0062) | | | | |
| Attend Community Events | | | | | -0.0016 | 0.0007 | | |
| | | | | | (0.0220) | (0.0154) | | |
| Subjective Integration | | | | | | | -0.0117 | -0.0218 |
| | | | | | | | (0.0182) | (0.0160) |
| Demographics Control | N | Y | Y | Y | N | Y | N | Y |
| Building Feature Control | N | Y | Y | Y | N | Y | N | Y |
| Time Fixed Effect | Y | Y | Y | Y | Y | Y | Y | Y |
| District Fixed Effect | Y | Y | Y | Y | Y | Y | Y | Y |
| First-Stage F-stats | | | 11.17 | 10.55 | | | | |
| Observations | 132 | 132 | 115 | 131 | 132 | 132 | 132 | 132 |
| R-squared | 0.258 | 0.702 | 0.687 | 0.692 | 0.257 | 0.698 | 0.260 | 0.704 |

Note: Panels A and B replicate the analyses in Table 3 using subsamples of Chinese and non-Chinese migrant renters, respective. Samples that did not rent a whole flat are excluded. Columns (1)-(2) report the OLS estimation result using immigration years as the independent variable. Column (3) reports the result using the growth rate of the foreign workforce as the main IV. Column (4) reports the result using the growth rate of the foreign population as the alternative IV. In Columns (3) and (4), the numbers of observations of the IV estimations are smaller due to missing information on the foreign workforce and population in the early years. Columns (5)-(6) report OLS estimation results using frequency of attendance at community activities as the independent variables, and Columns (7)-(8) report OLS estimation results using the subjective score of integration as the independent variables. Unreported demographic controls are same as in Table 2, which include age, gender, marital status, income, race, with child dummy, and ever moved home dummy. All variables are defined in Online Appendix Table 1. Standard errors are clustered at building level for OLS estimations and at arrival-year level for IV estimations. Robust standard errors in parentheses. *** p<0.01, ** p<0.05, * p<0.1

## Appendix B. Supplementary Tables

**Appendix Figure 1: Location of the North West District**

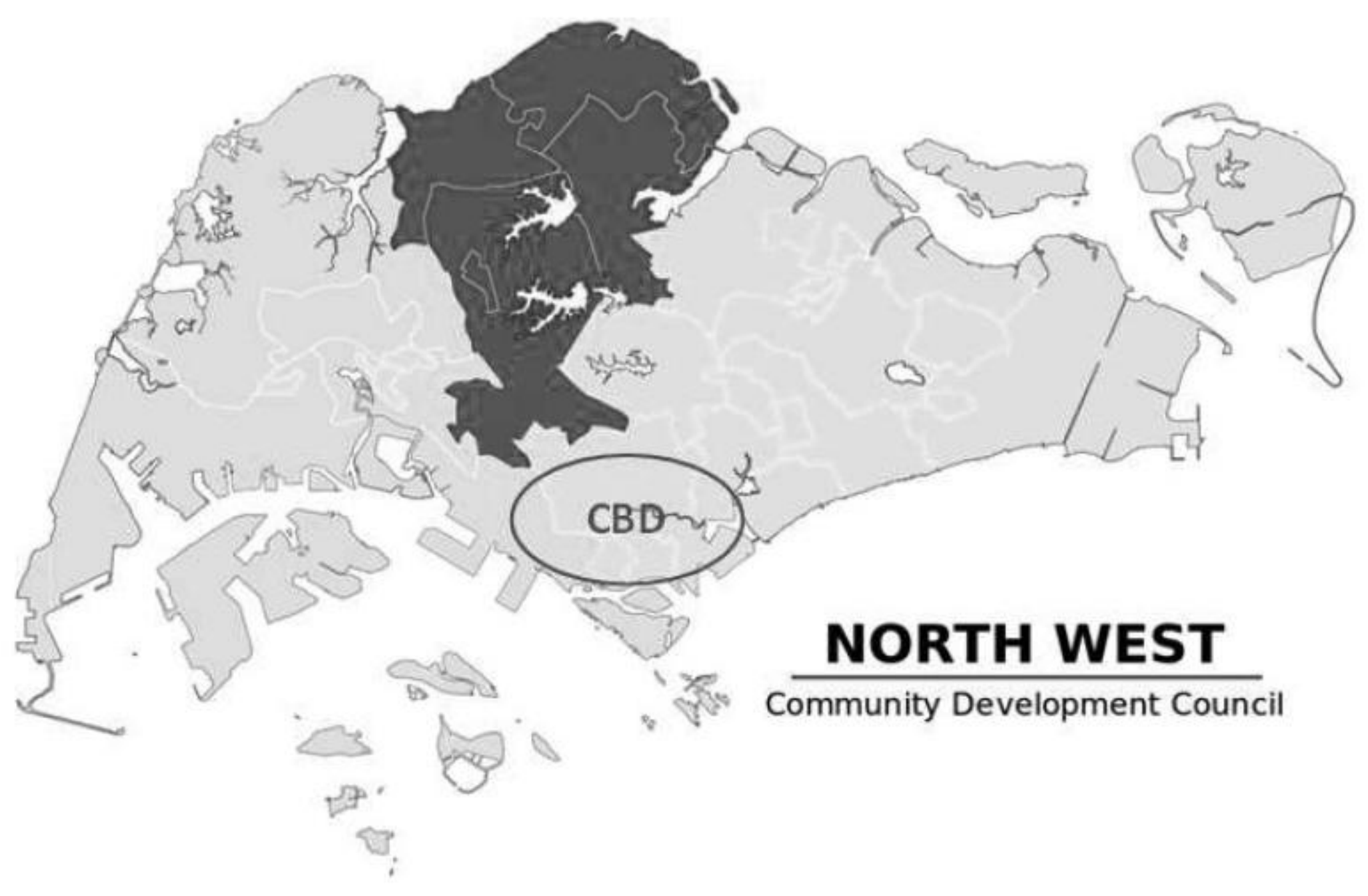

**Appendix Figure 2: Location of the Sampled Buildings**

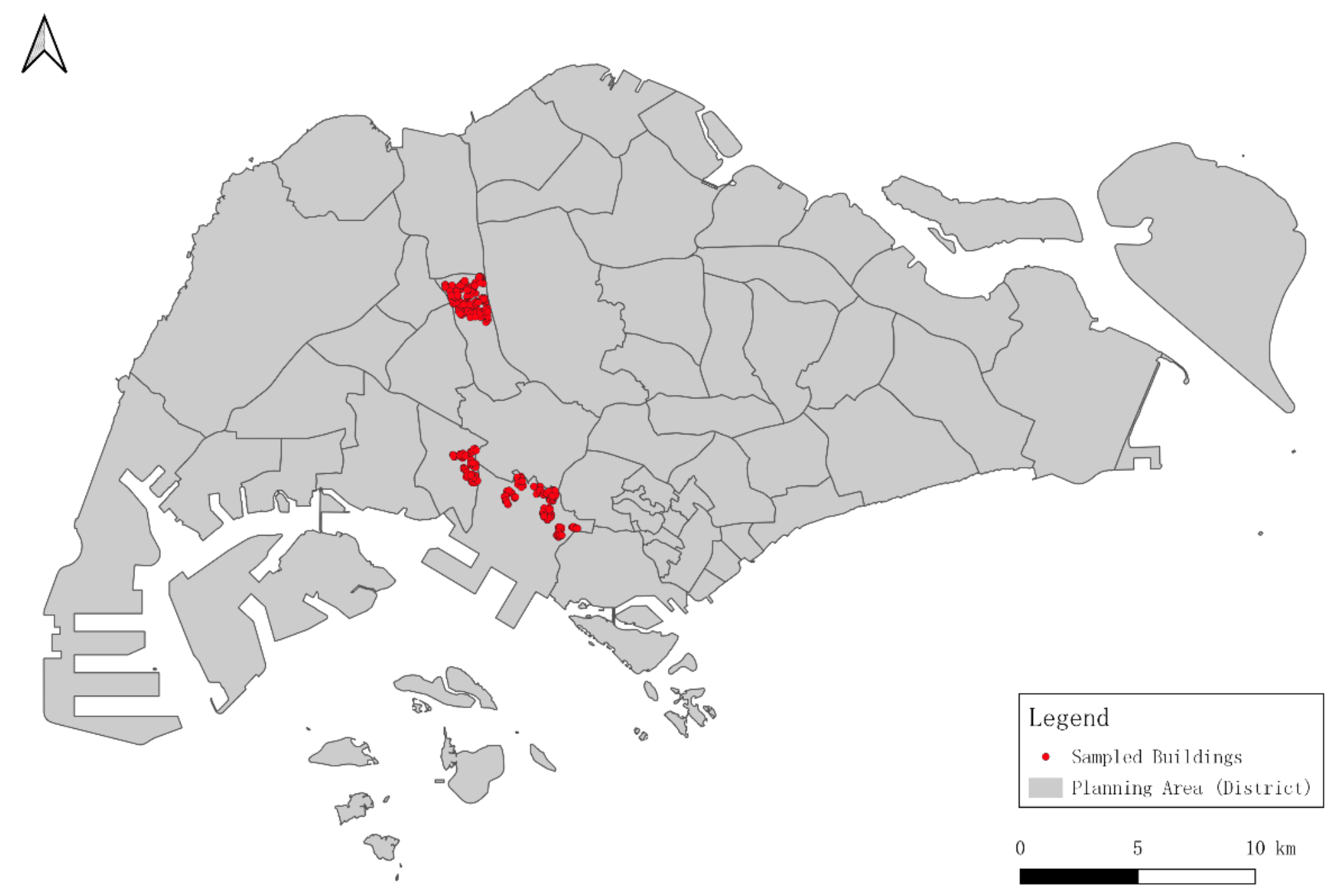

**Appendix Figure 3. Hedonic Price and Rent Index in Our Sample and in Singapore**

Note: The price and rent indices derived from our sample are estimated using a hedonic model (Rosen, 1974). The nationwide indices of Singapore are obtained from the Singapore Real Estate Exchange (SRX), which are also estimated using the hedonic method.

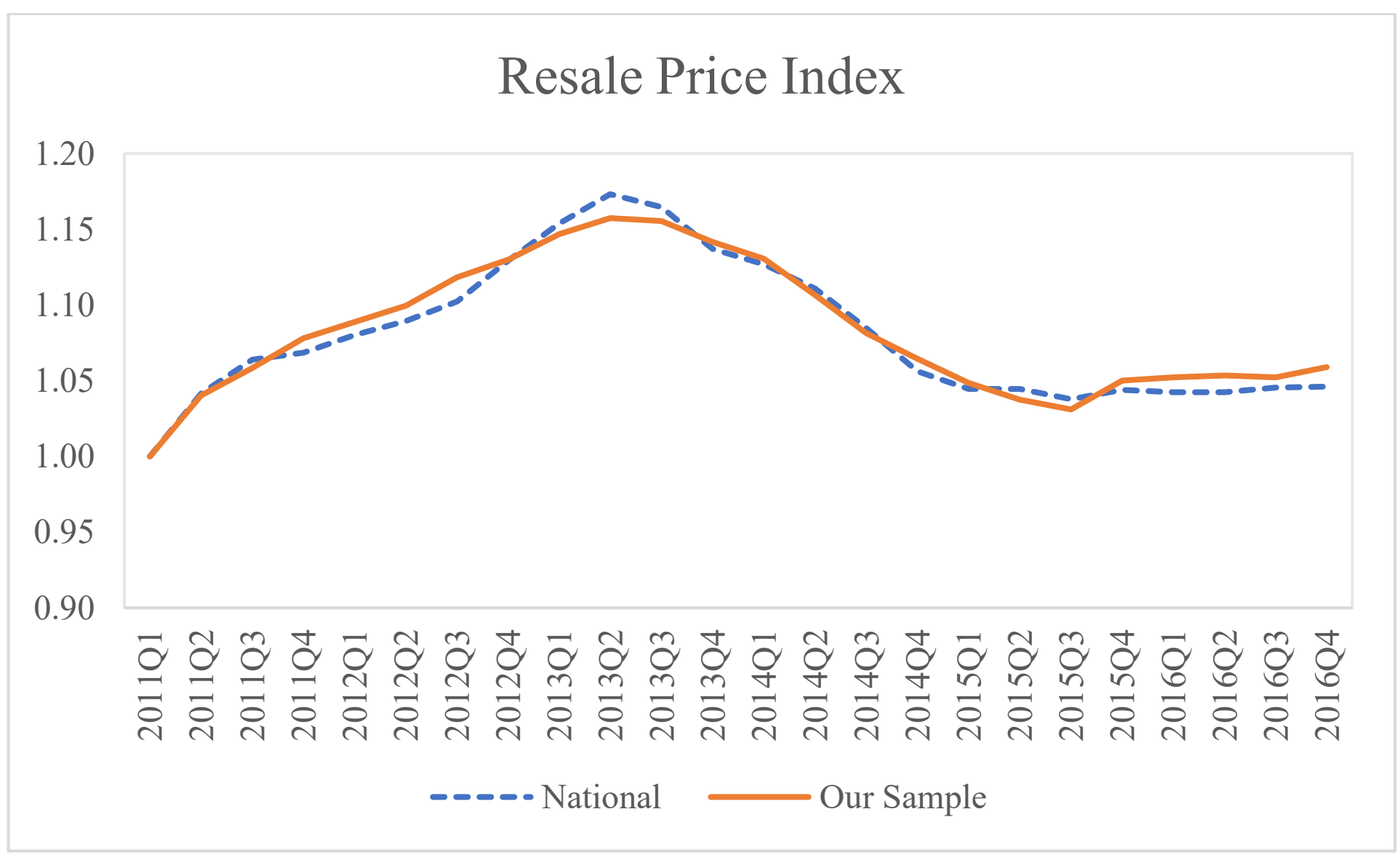

(a) Resale Price Index

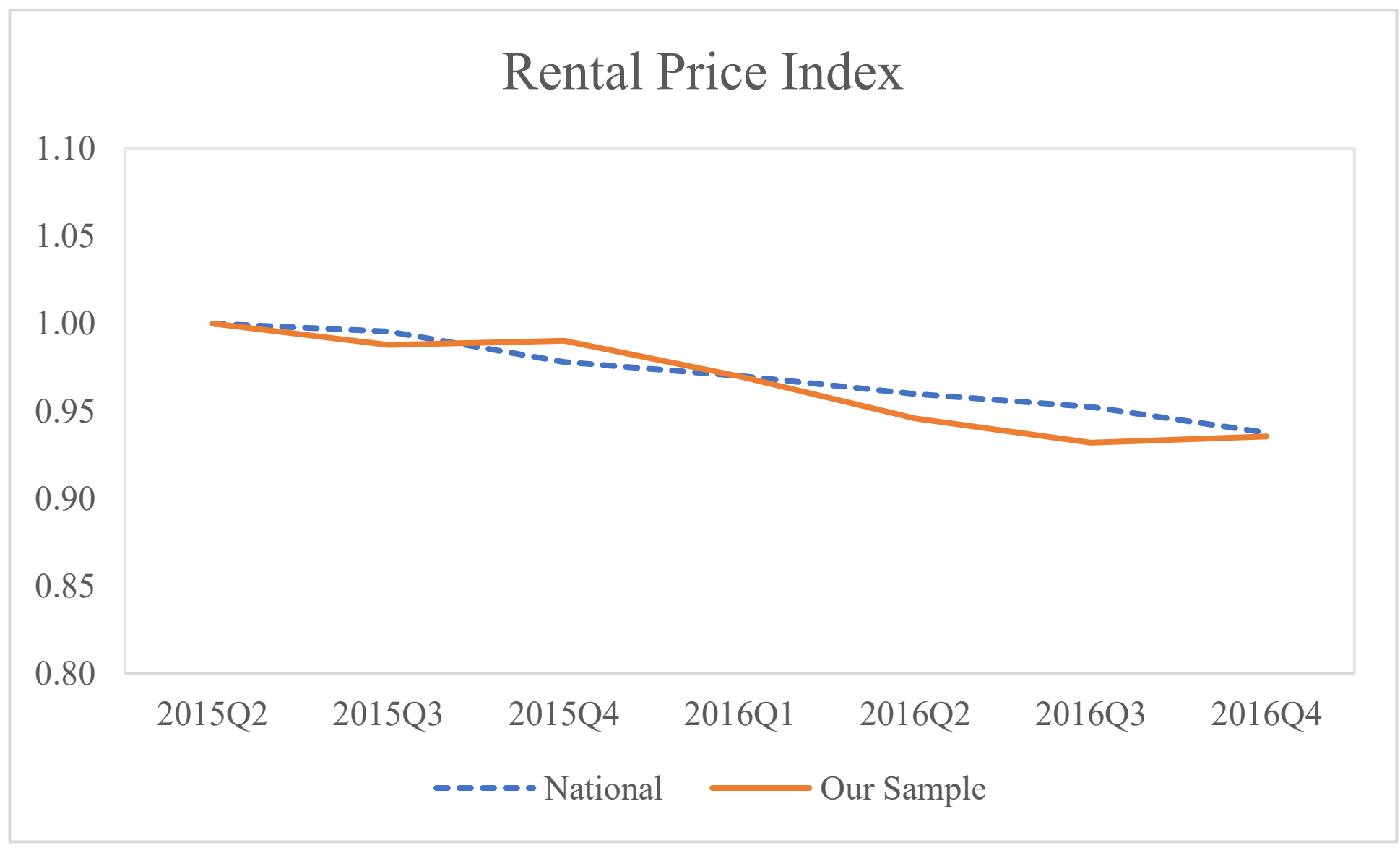

(b) Rental Price Index

## Appendix C. Details of Identification Strategy

Confounding factors, such as spatial polarisation of cities, or residential segregation by nationality or ethnicity, can contaminate the causal impact of social integration on immigrants' housing patterns. Singapore's public housing setting allows us to rule out major confounding factors at the city, neighbourhood, and individual levels, which would otherwise correlate with social integration and affect immigrants' housing behaviours simultaneously.

First, at the city level, spatial polarisation induced by economic development can concurrently affect immigrants' social integration and housing behaviours (Sassen, 1991). For example, job decentralisation and the suburbanisation of industries in receiving cities, along with economic growth and restructuring, induce spatial mismatches between the demand and supply of housing and employment, affecting immigrants' integration, housing choice, and commuting distance (Wilson, 1987; Holzer, 1991; Ihlanfeldt & Sjoquist, 1998; Dawkins, 2009). Such trends have been observed in global cities such as New York and Los Angeles, where there are large numbers of low-wage immigrants (Hamnett, 2021).

Nevertheless, the institutional features of Singapore largely shut off the alternative channels of spatial polarisation. Specifically, Singapore is a 'well-managed city' (Glaeser, 2012). Spatial mismatch between demand and supply of housing and employment across space is minimised through meticulous urban planning and development: First, the supply of public housing is geographically widespread, and the housing types are generally homogeneous across the city-state. Second, unlike most western cities, the city state of Singapore is compact and does not have distinct suburbs, so there is no suburbanisation of industries. Third, the well-developed public transport system of Singapore further mitigates the potential spatial mismatch between employment and housing (Zhu, 2018).

Second, at the neighbourhood level, residential segregation may exist in many global cities, influencing immigrants' housing outcomes. Residential segregation refers to the spatial separation and sorting of social groups into different neighbourhoods (Peach, 1975; van Kempen & Ozuekren, 1998; Phillips, 2010), and is considered to be a major barrier for immigrants to socially integrate into the receiving city (Friedrichs et al., 2003; Maloutas, 2004; Galster, 2007; Candipan et al., 2021). As summarised by Bolt et al. (2010), the process of integration not only depends on "*the characteristics of the immigrants, but also the reactions of the institutions and the population of the receiving society*". Conversely, a lack of integration is also seen as a driving force of forming residential segregation (Kalra & Kapoor, 2009; Simpson & Peach, 2009).

Once again, we use the unique institutional setting in Singapore and carefully design our empirical strategy to rule out the confounding effect of residential segregation by ethnicity or nationality, to isolate the causal impact of social integration on housing behaviours. As discussed in Section 2.2, given the strict quota restrictions across public housing buildings/estates, Singapore does not have evident residential segregation by immigration status or ethnicity in its public housing, which accommodates over 80% of the population.

Third, at the individual level, demographic and socioeconomic backgrounds of immigrants can influence housing behaviours. For example, individual life-stage changes (e.g., marriage or childbirth) during post-migration years can have significant effects in determining housing choices and residential mobility (Clark & Onaka, 1983). Occupation type and income level of immigrants can also impact the importance of proximity to their workplace. To rule out the effects of these individual-level confounding factors, we collect a comprehensive list of demographic and socioeconomic information, and explicitly control for these factors in our multivariate regression analyses to isolate the impact of social integration on housing choices.

Also, political preference can influence individuals' housing patterns. For instance, in some western countries, immigrants may have spatial sorting due to political preferences (Gimpel & Hui, 2015).

However, neighbourhood selection based on political concerns in not commonplace in Singapore, as there has been only one dominant political party since 1965. There is also little participation in political activities among immigrants.

Another concern at the individual level is that changes in immigration status over post-migration years can impact immigrants' housing behaviours. Nevertheless, in our empirical designs, we only compare market participants with the same immigration statuses (i.e., ***within*** public housing renters, who are almost all immigrants; or ***within*** public housing homeowners, who are citizens, no matter naturalised or locally born). As the immigration status within each subgroup is homogenous, the differences in housing outcomes ***within*** each subgroup are not caused by differences in immigration status.

Finally, we further control for time fixed effects in our regression models, which help eliminate the impact of unobserved temporal variations, such as national political events, if any, that may affect immigrants' housing behaviours. In addition to time fixed effects, we also control for district (i.e., HDB town) fixed effects in our models. They capture the remaining spatial variations in the outcome variables which may not be taken by existing control variables, such as distances to amenities at the district level.

**Appendix References:**

Bolt, G., Özüekren, A. S., & Phillips, D. (2010). Linking Integration and Residential Segregation. *Journal of Ethnic and Migration Studies, 36*(2), 169-186.

Candipan, J., Phillips, N. E., & Small, M. (2021). From residence to movement: The nature of racial segregation in everyday urban mobility. *Urban Studies, 58*(15), 3095-3117.

Clark, W. A., & Onaka, J. L. (1983). Life Cycle and Housing Adjustment as Explanations of Residential Mobility. *Urban Studies, 20*(1), 47-57.

Glaeser, E. (2012). *Triumph of the City: How Our Greatest Invention Makes Us Richer, Smarter, Greener, Healthier, and Happier.* Penguin Books.

Dawkins, C. J. (2009). Exploring Recent Trends in Immigrant Suburbanization. *Cityscape, 11*(3), 81-97.

Friedrichs, J., Galster, G., & Musterd, S. (2003). Neighbourhood effects on social opportunities: the European and American research and policy context. *Housing Studies, 18*(6), 797-806.

Galster, G. (2007). Neighbourhood Social Mix as a Goal of Housing Policy: A Theoretical Analysis. *European Journal of Housing Policy, 7*(1), 19-43.

Gimpel, J. G., & Hui, I. S. (2015). Seeking Politically Compatible Neighbors? The Role of Neighborhood Political Composition in Residential Sorting. *Political Geography, 48*, 130-142.

Hamnett, C. (2021). The changing social structure of global cities: Professionalisation, proletarianisation or polarisation. *Urban Studies, 58*(5), 1050-1066.

Holzer, H. J. (1991). The Spatial Mismatch Hypothesis: What Has the Evidence Shown? *Urban Studies, 28*(1), 105-122.

Ihlanfeldt, K., & Sjoquist, D. L. (1998). The spatial mismatch hypothesis: A review of recent studies and their implications for welfare reform. *Housing Policy Debate, 9*(4), 849-892.

Kalra, V. S., & Kapoor, N. (2009). Interrogating Segregation, Integration and the Community Cohesion Agenda. *Journal of Ethnic and Migration Studies, 35*(9), 1397-1415.

Maloutas, T. (2004). Segregation and Residential Mobility: Spatially Entrapped Social Mobility and Its Impact on Segregation in Athens. *European Urban and Regional Studies, 11*(3), 195-211.

Peach, C. (1975). *Urban Social Segregation.* London: Longman Publishing Group.

Phillips, D. (2010). Minority Ethnic Segregation, Integration and Citizenship: A European Perspective. *Journal of Ethnic and Migration Studies, 36*(2), 209-225.

Sassen, S. (1991). *The Global City: New York, London, Tokyo.* Princeton, NJ: Princeton University Press.

Simone, D., & Newbold, B. (2014). Housing Trajectories Across the Urban Hierarchy: Analysis of the Longitudinal Survey of Immigrants to Canada, 2001–2005. *Housing Studies, 29*(8), 1096-1116.

Simpson, L., & Peach, C. (2009). Measurement and Analysis of Segregation, Integration and Diversity. *Journal of Ethnic and Migration Studies, 35*(9), 1237-1380.

van Kempen, R., & Ozuekren, A. (1998). Ethnic Segregation in Cities: New Forms and Explanations in a Dynamic World. *Urban Studies, 35*(10), 1631-1656.

Wilson, W. J. (1987). *The Truly Disadvantaged: The Inner City, the Underclass, and Public Policy.* Chicago: University of Chicago Press.

Zhu, M. (2018). *Singapore's public transport system among best in the world: McKinsey report.* Retrieved from Channel News Asia: https://www.channelnewsasia.com/singapore/singapore-public-transport-system-among-best-in-the-world-805051